\documentclass[aos,preprint]{imsart}
\setattribute{journal}{name}{}
\usepackage[dvips]{graphicx}

\usepackage[tbtags]{amsmath}
\usepackage{amsbsy}
\usepackage{amssymb}
\usepackage{amsfonts}
\usepackage{amsthm}

\usepackage{graphicx}
\usepackage{url}
\usepackage{algorithmic}
\usepackage{algorithm}
\usepackage{balance}
\usepackage{rotating}
\usepackage{multirow}
\usepackage{subcaption}
\usepackage{fancyvrb}
\usepackage{latexsym}
\usepackage{verbatim}
\usepackage{ctable}

\newtheorem{theorem}{Theorem}

\newtheorem{lemma}{Lemma}
\newtheorem{corollary}{Corollary}

\newtheorem{definition}{Definition}

\newtheorem{condition}{Condition}

{\begin{list}               %    with flush left bullets.
    {$\bullet$ \hfill}{
        \setlength{\leftmargin}{\parindent}
        \setlength{\parsep}{0.04\baselineskip}
        \setlength{\itemsep}{0.5\parsep}
        \setlength{\labelwidth}{\leftmargin}
        \setlength{\labelsep}{0em}}
    }
{\end{list}}

\providecommand{\eref}[1]{\eqref{#1}}  % call \eqref from amstex
\providecommand{\cref}[1]{Chapter~\ref{#1}}

\providecommand{\fref}[1]{Figure~\ref{#1}}

\providecommand{\thref}[1]{Theorem~\ref{#1}}

\providecommand{\R}{\ensuremath{\mathbb{R}}}

\providecommand{\E}{\ensuremath{\mathbb{E}}}
\providecommand{\N}{\ensuremath{\mathbb{N}}}

\providecommand{\Pb}{\ensuremath{\mathbb{P}}}

\renewcommand{\vec}[1]{\ensuremath{\boldsymbol{#1}}}

\providecommand{\calO}{\mathcal{O}}
\providecommand{\calL}{\mathcal{L}}

\providecommand{\calE}{\mathcal{E}}

\providecommand{\vhat}{\widehat{v}}
\providecommand{\uhat}{\widehat{u}}
\providecommand{\zhat}{\widehat{z}}

\providecommand{\what}{\widehat{w}}

\providecommand{\rhat}{\widehat{r}}

\providecommand{\Ahat}{\widehat{A}}

\providecommand{\Hhat}{\widehat{H}}

\providecommand{\sigmahat}{\widehat{\sigma}}

\providecommand{\vone}{\vec{1}}

\providecommand{\Var}{\mathrm{Var}}

\newcommand{\defequal}{\mathop{\overset{\mbox{\tiny{def}}}{=}}}
\newcommand{\subjectto}{\mathop{\mathrm{subject\, to}}}
\newcommand{\argmin}[1]{\mathop{\underset{#1}{\mbox{argmin}}}}
\newcommand{\argmax}[1]{\mathop{\underset{#1}{\mbox{argmax}}}}

\providecommand{\MSE}{\mathrm{MSE}}

\graphicspath{{pix_2/}}

\RequirePackage[OT1]{fontenc}
\RequirePackage{natbib}
 \bibpunct{(}{)}{;}{a}{,}{,}
\RequirePackage[colorlinks,citecolor=blue,urlcolor=blue]{hyperref}

\startlocaldefs
\numberwithin{equation}{section}
\theoremstyle{plain}

\endlocaldefs

\begin{document}

\begin{frontmatter}
\title{A Consistent Histogram Estimator for Exchangeable Graph Models}
\runtitle{A Consistent Histogram Estimator for Exchangeable Graph Models}

\begin{aug}
\author{\fnms{Stanley H.}     \snm{Chan} \thanksref{t1,m1} \ead[label=e1]{schan@seas.harvard.edu}},
and
\author{\fnms{Edoardo M.}     \snm{Airoldi} \thanksref{t2,m1} \ead[label=e2]{airoldi@fas.harvard.edu}}
\thankstext{t1}{SHC is partially supported by a Croucher Foundation Post-Doctoral Research Fellowship.}
\thankstext{t2}{EMA is partially supported by NSF CAREER award IIS-1149662, ARO MURI award W911 NF-11-1-0036, and an Alfred P. Sloan Research Fellowship.}
\runauthor{Chan \& Airoldi}

\affiliation{Harvard University\thanksmark{m1}}

\address{Department of Statistics,\\
1 Oxford Street, \\
Cambridge, MA 02138.
\printead{e1,e2}
\phantom{E-mail: \ }}

%\address{Address of the Third author\\
%Usually a few lines long\\
%Usually a few lines long\\
%\printead{e3}\\
%\printead{u1}}
\end{aug}

\begin{abstract}
Exchangeable graph models (ExGM) subsume a number of popular network models. The mathematical object that characterizes an ExGM is  termed a {\it graphon}. Finding scalable estimators of graphons, provably consistent, remains an open issue. In this paper, we propose a histogram estimator of a graphon that is provably consistent and numerically efficient. The proposed estimator is based on a sorting-and-smoothing (SAS) algorithm, which first sorts the empirical degree of a graph, then smooths the sorted graph using total variation minimization. The consistency of the SAS algorithm is proved by leveraging  sparsity concepts from compressed sensing.
\end{abstract}

\begin{keyword}[class=MSC]
\kwd[Primary]{62G05}
\kwd{90B15}
\kwd[; secondary ]{05C62}
\end{keyword}

\begin{keyword}
\kwd{Graphon}
\kwd{Non parametric network model}
\kwd{Total variation}
\kwd{Exchangeable random graph model}
\kwd{Network histogram}
\end{keyword}

\end{frontmatter}

\section{Introduction}
Developing statistical models for network data has been a growing research area in statistics and machine learning over the past decade~\citep{Goldenberg_Zheng_Fienberg_2009, Kolaczyk_2009, Airoldi_Bai_Carley_2011}. Among many models, the \emph{parametric} families have been the major focus in the literature because of their simplicity and analytic tractability. Popular examples of these parametric models include the exponential random graph model \cite{Wasserman_2005,Hunter_Handcock_2006}, the stochastic blockmodel \cite{Nowicki_Snijders_2001}, the mixed membership model \cite{Airoldi_Blei_Fienberg_2008}, the latent space model \cite{Hoff_Raftery_Handcock_2002}, the graphlet \cite{Azari_Airoldi_2012} and many others. However, as the complexities of the networks increase, it becomes increasingly more challenging to fit the data using a particular parametric model.

\subsection{Non-parametric representation of a graph}
In this paper, we consider a non-parametric perspective of modeling network data using the exchangeable graph models (ExGM). The notion of exchangeability is due to de Finetti, later generalized by Aldous~\cite{Aldous_1981}, Hoover \cite{Hoover_1979} and Kallenberg \cite{Kallenberg_2005}. A connection between popular parametric models and  exchangeable graph models has been recently made \cite{Hoff_2008,Bickel_Chen_2009}.

The \emph{non-parametric} (limit) object that characterizes an ExGM is often termed a \emph{graphon}. As we will define formally in Section 2, a graphon is a $2$-dimensional continuous function on $[0,1]^2 \rightarrow [0,1]$ that generates random graphs. Since a graphon is a model for network data, any model based inference, prediction and hypothesis testing can be performed using a graphon \cite{Lloyd_Orbanz_Ghahramani_2012}. For instance, when comparing networks that span different sample sizes, graphons provide a natural solution: If two samples of a network are generated from the same ExGM, they should have the same graphon, and hence, comparing two networks can be done by comparing two graphons.

In this paper, we propose an efficient graphon estimator based on 2D histograms. The challenge of the problem is two-fold. First, since graphons are unique up to measure-preserving transformations, it is important to identify the conditions under which graphons can be uniquely recovered \citep[e.g., see][]{Yang:2013fk}. Second, it is desirable for a graphon estimator to be provably consistent.

\subsection{Related work}
Previous methods of graphon estimation algorithms can be classified into two categories as follows.

The first category is to perform graphon estimation conditioned on the node arrangement. When the node arrangement is conditioned on, we can bypass the difficult problem of identifying a canonical representation of the graphon. For example, the universal singular value thresholding \cite{Chatterjee_2012} and the matrix completion \cite{Keshavan_Montanari_Oh_2010} seek low-rank structures of the adjacency matrix, whereas the stochastic blockmodel approximation \cite{Airoldi_Costa_Chan_2013,Chan_Costa_Airoldi_2013_globalsip} groups similar nodes to form community structures. However, since the estimations are conditioned on the node arrangement, the resulting graphons are not canonical.

Different from the first category, the second category of methods estimate \emph{canonical} graphons. In \cite{Bickel_Chen_Levina_2011}, the authors proposed a method of moments which is theoretically consistent for that purpose. However, the method requires knowledge of \emph{all} wheels of the network, and hence is computationally infeasible. Choi et al. \cite{Choi_Wolfe_Airoldi_2012, Choi_Wolfe_2012} attempted the problem by a clustering approach, but they stopped at the clustering step without actually estimating the graphon. In \cite{Lloyd_Orbanz_Ghahramani_2012}, Lloyd et al. considered a Bayesian approach to estimate a graphon. However, the MCMC sampling process of the algorithm is computationally intensive. Moreover, there is no consistency guarantee of the estimator. More recently, other groups have begun exploring alternative approaches \cite{Wolfe_Olhede_2013, Tang_Sussman_Priebe_2013, Latouche_Robin_2013, Olhede_Wolfe_2013}. Yet, none of these methods are \emph{both} consistent and computationally efficient.

\subsection{Contributions}
In this paper, we propose a histogram approach to estimate graphons. Our method, called the Sorting-And-Smoothing (SAS) algorithm, consists of two steps. In the first step, we sort the empirical degrees and rearrange the nodes of the graph for a canonical ordering. In the second step, we compute the histogram of the sorted graph and smooth the histogram by a total variation minimization. Details of the SAS algorithm are presented in Section 3.

The estimator returned by the SAS algorithm is consistent. The consistency proof leverages the sparsity concepts from compressed sensing. In particular, we show, in Theorem 3, that if the true graphon satisfies some Lipschitz conditions and has sparse gradients, then the mean squared error (MSE) of the estimator is $\calO( (\log n) / n)$, where $n$ is the size of the network. Discussion of the consistency is presented in Section 4.

We test the SAS algorithm on both simulation data and real data (Section 5). The experiment of using the simulation data indicates that the SAS algorithm is superior to, both in terms of estimation quality and speed, several existing methods. Applying the SAS algorithm to real data, we estimate graphons of two large-scale social networks and reveal some structures. These results provide an alternative way of analyzing large-scale network data.

\section{Graphons and identifiability}
The purpose of this section is to introduce the concepts of a graphon and discuss the conditions under which a graphon can be uniquely identified.

\subsection{Definition of a graphon}
We let $G$ be the adjacency matrix of a graph with the $(i,j)$th entry denoted by $G_{ij} \in \{0,1\}$. For an infinitely sized graph $G$, we say that $G$ is exchangeable if it satisfies the following definition.
\begin{definition}
\label{def:exchangeable}
An infinite random array $G = (G_{ij})_{i,j \in \N}$ is exchangeable if
\begin{equation}
(G_{ij}) \overset{d}{=} (G_{\sigma(i)\sigma(j)}),
\end{equation}
for any permutation $\sigma$.
\end{definition}
Definition \ref{def:exchangeable} is also known as the \emph{joint} exchangeability, because the permutation is applied to both rows and columns simultaneously \cite{Orbanz_Roy_2013}.

We refer to all random graph models that satisfy exchangeability as exchangeable graph models (ExGM). A useful characterization of an ExGM is given by the Aldous-Hoover theorem.
\begin{theorem}[Aldous-Hoover]
\label{thm:aldous-hoover}
An infinite random array $(G_{ij})_{i,j \in \N}$ is exchangeable if and only if there is a random measurable function $F: [0,1]^3 \rightarrow \{0,1\}$ such that
\begin{equation}
(G_{ij}) \overset{d}{=} (F(U_i,U_j,U_{ij})),
\end{equation}
where $(U_i)_{i \in \N}$ and $(U_{ij})_{i,j \in \N}$ are sequences of i.i.d. Uniform$[0,1]$ random variables.
\end{theorem}

The function $F$ in \thref{thm:aldous-hoover} defines a \emph{graphon}:
\begin{definition}[Graphon]
A graphon $w$ is a symmetric measurable function $w: [0,1]^2 \rightarrow [0,1]$ such that
\begin{equation}
F(U_i,U_j,U_{ij}) =
\begin{cases}
1, &\mbox{ if } U_{ij} < w(U_i,U_j)\\
0, &\mbox{ otherwise, }
\end{cases}
\label{eq:simplified eq Aldous Hoover}
\end{equation}
where $(U_i)_{i \in \N}$ and $(U_{ij})_{i,j \in \N}$ are sequences of i.i.d. Uniform$[0,1]$ random variables.
\end{definition}

Equivalently, \eref{eq:simplified eq Aldous Hoover} can be expressed as the following two-stage sampling scheme:
\begin{equation}
\begin{array}{ll}
U_i                   &\overset{iid}{\sim} \mbox{Uniform}[0,1],\\
G_{ij} \mid U_i, U_j  &\sim \mbox{Bernoulli}(w(U_i, U_j)).
\end{array}
\label{eq:sampling}
\end{equation}
Therefore, a finite sized network generated from a graphon can be regarded as a finite sample drawn according to \eref{eq:sampling}.

\subsection{Identifiability of a graphon}
To understand the identifiability issue of a graphon, it is important to discuss measure preserving transformations.
\begin{definition}[Measure Preserving Transformation]
\label{def:MPT}
A transformation $\varphi: [0,1] \rightarrow [0,1]$ is measure-preserving w.r.t. a measure $\mu$ if it is measurable, and for all $A \in [0,1]$,
\begin{equation}
\mu(\varphi^{-1}(A)) = \mu(A).
\label{eq:MPT}
\end{equation}
\end{definition}
For example, if $\varphi$ is a measure preserving transformation and $U \sim \mathrm{Uniform}[0,1]$, then $\varphi(U)$ is also distributed uniformly on $[0,1]$. Similarly, if $\varphi$ is a measure preserving transformation, then the graphon
\begin{equation*}
w'(u,v) \defequal w(\varphi(u), \varphi(v))
\end{equation*}
defines the same ExGM as $w$ because there exists a transformation such that $w$ and $w'$ are identical.

The identifiability issue of a graphon arises because the converse of Definition~\ref{def:MPT} is not true in general: If $w$ and $w'$ define the same ExGM, there may not exist a measure preserving transformation $\varphi'$ such that $w(u,v) = w'(\varphi'(u),\varphi'(v))$ \cite{Diaconis_Janson_2007}. For example, the functions $w(u,v) = uv$ and $w'(u,v) = (2 u \mod 1)(2 v \mod 1)$ define the same ExGM, but there is no $\varphi'$ such that $w(u,v) = w'(\varphi'(u),\varphi'(v))$.

A formal statement of the above observation is given by the following theorem, which says that we need to find a \emph{pair} of measure-preserving transformations $\varphi$ and $\varphi'$ in order to show that $w$ is unique.

\begin{figure}[h]
\centering
\begin{tabular}{cc}
\includegraphics[width=0.25\linewidth]{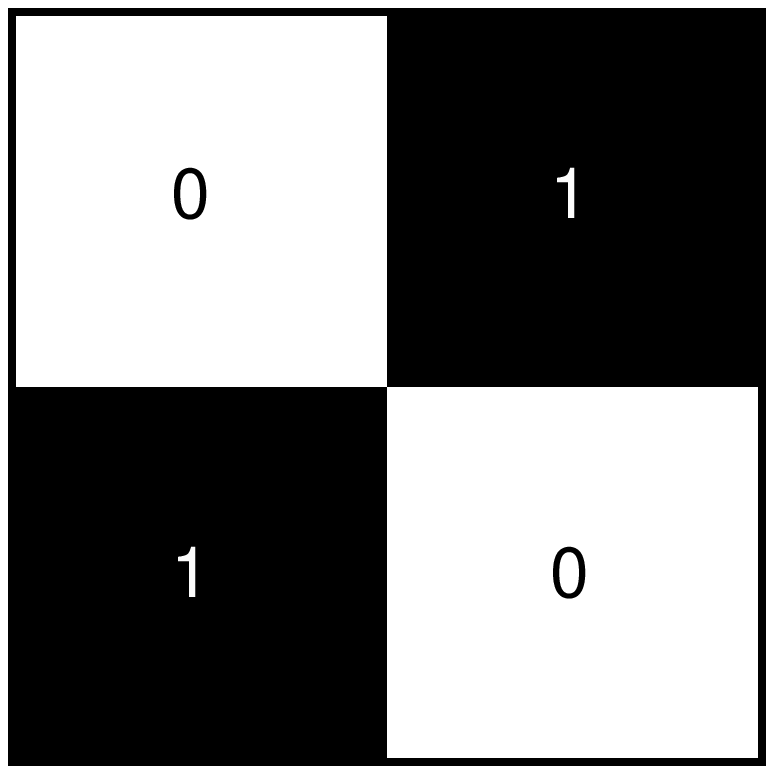}&
\includegraphics[width=0.25\linewidth]{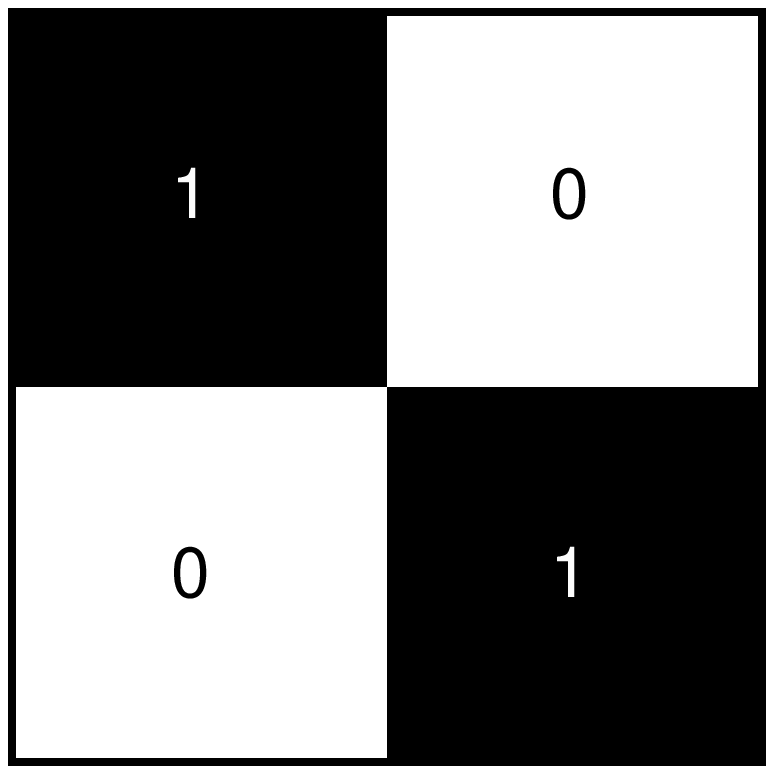}\\
$w$ & $w'$
\end{tabular}
\caption{Example of a pair of twin graphons: $w$ and $w'$ are not identifiable if we randomly permute their columns and rows.}
\label{fig:twin}
\end{figure}

\begin{theorem}[\cite{Diaconis_Janson_2007}, Thm. 7.1]
\label{thm:Diaconis}
Let $w$ and $w'$ be two graphons. Then $\delta_{\square}(w,w') = 0$ if and only if there exist measure-preserving transformations $\varphi$ and $\varphi':[0,1] \rightarrow [0,1]$ such that
\begin{equation}
w(\varphi(u),\varphi(v)) = w'(\varphi'(u),\varphi'(v)),
\end{equation}
where the distance $\delta_{\square}(w,w')$ is the cut-norm defined by \cite{Lovasz_Szegedy_2006}.
\end{theorem}
A consequence of \thref{thm:Diaconis} is the notion of twin-free:
\begin{definition}[Twin-free \cite{Borgs_Chayes_Lovasz_2010}]
A graphon $w$ is called \emph{twin-free} if for any $u_1$ and $u_2 \in [0,1]$, $w(u_1, v) \not= w(u_2, v)$ for almost all $v \in [0,1]$.
\end{definition}
Essentially, the twin-free condition excludes the cases where two graphons can be made identical by row and column permutations. For example, the pair shown in \fref{fig:twin} are twin, and hence they are not identifiable.

The twin-free condition is necessary but not sufficient for identifying a unique graphon when we marginalize a graphon \cite{Orbanz_Roy_2013}:
\begin{equation*}
g(u) \defequal \int_0^1 w(u,v) dv.
\end{equation*}
For example, if we consider $w$ and $w'$ in \fref{fig:twin}, and a graphon $w''(u,v) = 1/2$, then $w''$ is twin-free but $g(u) = g'(u) = g''(u)$, where $g$, $g'$ and $g''$ are marginalizations of $w$, $w'$ and $w''$, respectively.

The necessary and sufficient condition for a graphon to be identifiable is to require strict monotonicity of degrees \cite{Bickel_Chen_2009,Yang:2013fk}.
\begin{condition}[Strict Monotonicity of Degree]
A graphon $w$ has a unique representation if and only if there exists $w^{can}$ such that
\begin{equation*}
g^{can}(u) \defequal \int_0^1 w^{can}(u,v) dv
\end{equation*}
is strictly increasing (or decreasing). The graphon $w^{can}$ is called the canonical representation of $w$.
\end{condition}
It is evident that the strict monotonicity condition implies twin-free, but not vice versa. In addition, if we let $U \sim \mbox{Uniform}[0,1]$, then strict monotonicity implies that $g^{can}(U)$ is absolutely continuous.

In the rest of the paper we assume that all graphons of interests satisfy the strict monotonicity condition. For notational simplicity, we drop the superscript $(\cdot)^{can}$, and denote $w$ as the canonical representation.

\section{Proposed SAS Algorithm}
The intuition of the proposed SAS algorithm is based on the following idea: As the size of a graph grows, the (sorted) empirical degree should converge to the ideal (canonical) degree distribution. Therefore, if we can sort the empirical degree of a given graph, then by applying suitable smoothing algorithms we can find an estimate of the canonical graphon.

\begin{figure}[t]
\centering
\begin{tabular}{cc}
\includegraphics[height=3cm]{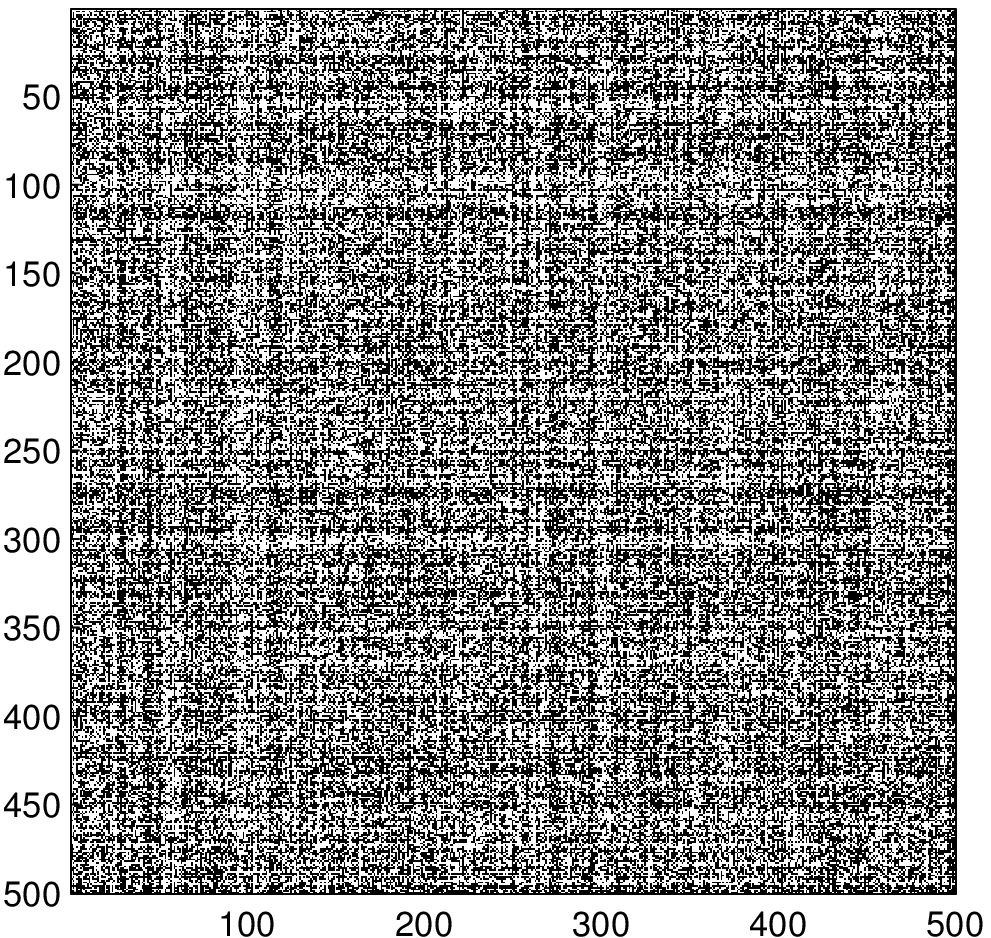}&
\includegraphics[height=3cm]{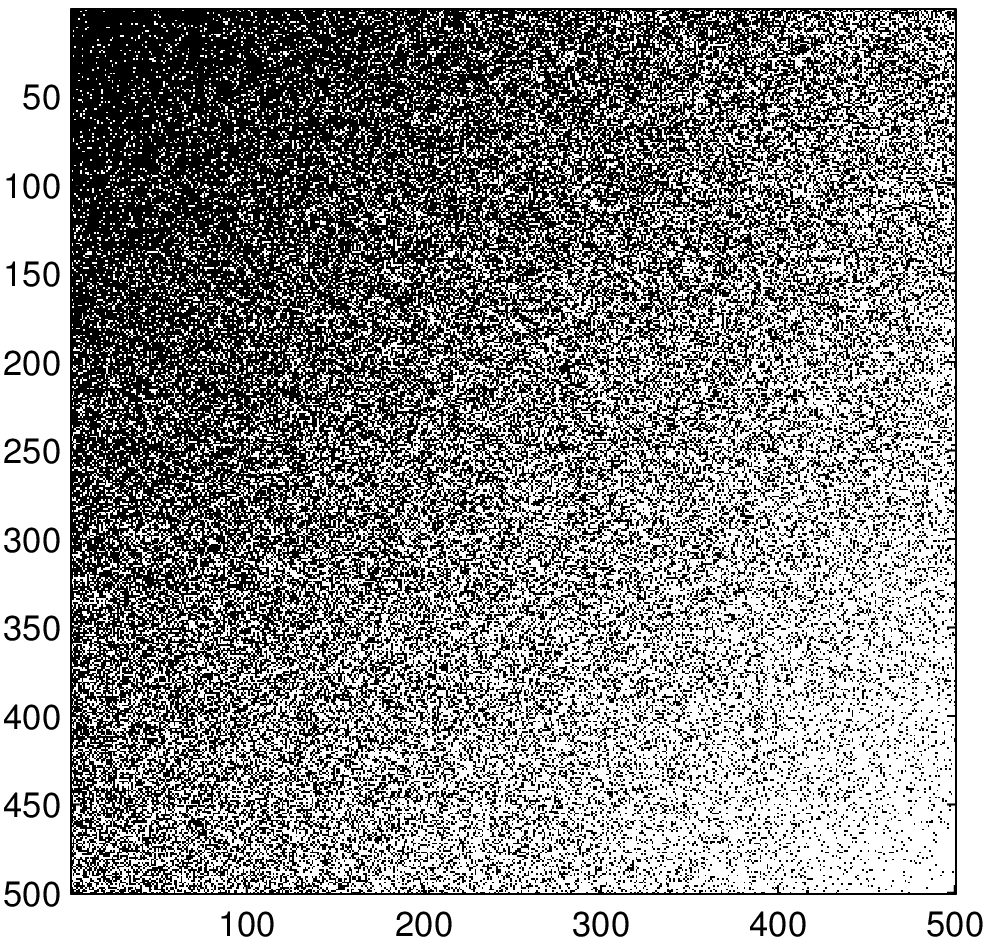}\\
Observed graph  & Sorted graph \\
$G \in \{0,1\}^{n \times n}$ & $\Ahat \in \{0,1\}^{n \times n}$ \\
\includegraphics[height=3cm]{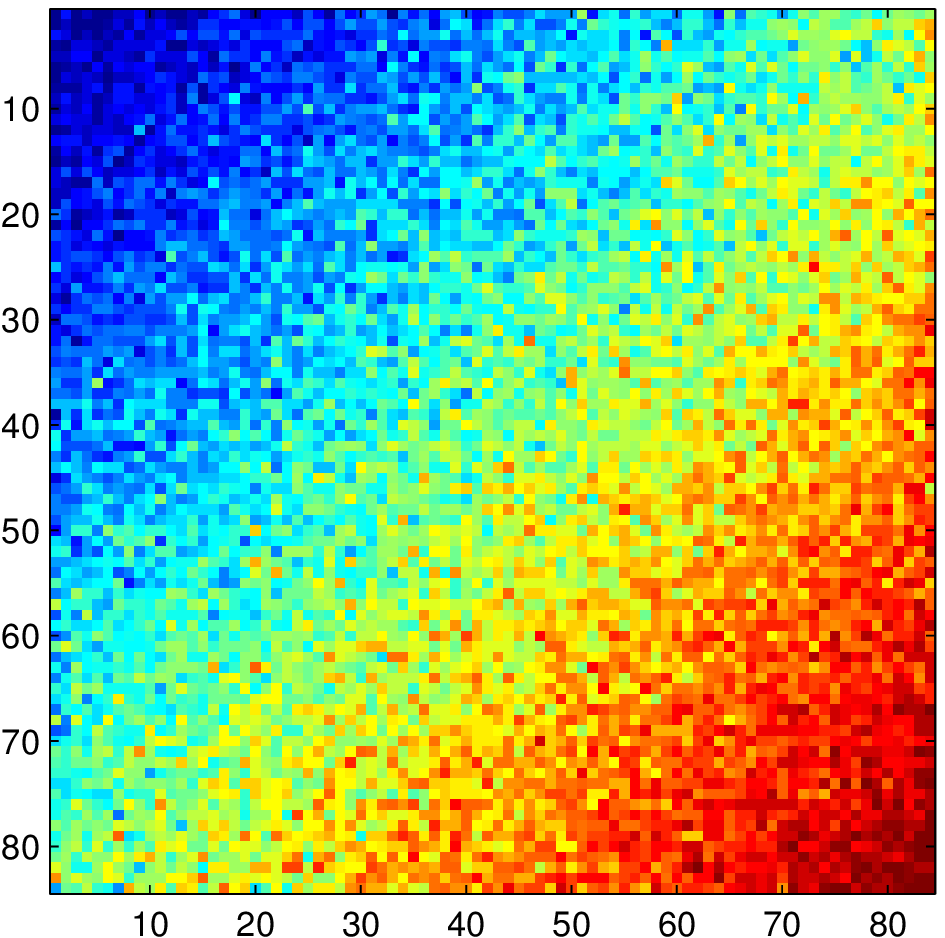}&
\includegraphics[height=3cm]{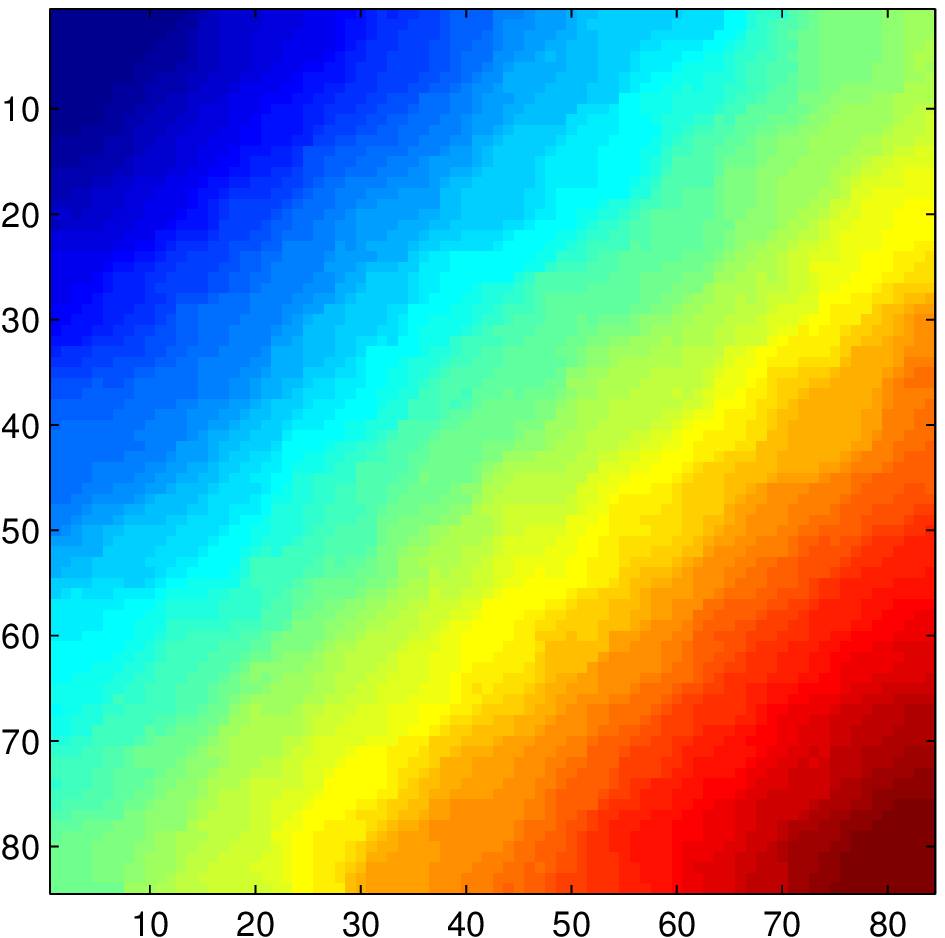}\\
Local histogram & Estimated graphon \\
$\Hhat \in [0,1]^{k \times k}$ & $\what^{tv} \in [0,1]^{k \times k}$
\end{tabular}
\caption{Illustration of the SAS algorithm. Given an observed graph $G$, we first sort $G$ using the empirical degrees to get $\Ahat$. Then, a local histogram $\Hhat$ is computed and a total variation minimization is used to determine an estimate $\what^{tv}$.}
\label{fig:illustration}
\end{figure}

Following this intuition, we propose a two-stage algorithm. In the first stage, we sort the rows and columns of $G$ to obtain a sorted graph $\Ahat$ according to the empirical degree. In the second stage, we compute a histogram $\Hhat$ of $\Ahat$, and apply a total variation minimization to find an estimate $\what^{tv}$. An illustration of the SAS algorithm is shown in \fref{fig:illustration}, and a pseudo code is shown in Algorithm~\ref{alg:sas}.

\begin{algorithm}[t]
\caption{Sort and Smooth (SAS) algorithm}
\begin{algorithmic}
\STATE Input: An $n \times n$ graph $G$.
\STATE Output: An $n \times n$ estimate of the graphon $\what^{est}$.
\STATE \textbf{Step 1 - Sorting}
\STATE \quad Compute empirical degree distribution $d_i = \sum_{j=1}^n G_{ij}$
\STATE \quad Sort the degree distribution and determine the corresponding permutation $\sigmahat$.
\STATE \quad Form a sorted graph $\Ahat_{ij} = G_{\sigmahat(i)\sigmahat(j)}$.
\STATE \textbf{Step 2 - Smoothing}
\STATE \quad Compute histogram $$\Hhat_{ij} = \frac{1}{h^2} \sum_{i_1=1}^h \sum_{j_1=1}^h \Ahat_{ih+i_1,jh+j_1},$$ for some binwidth $h$.
\STATE \quad Solve total variation minimization
\begin{equation*}
\what^{tv} = \argmin{\widehat{r}} \;\; \| \widehat{r} \|_{TV}, \quad\quad \subjectto \| \widehat{r} - \Hhat \|_2 \le \varepsilon
\end{equation*}
for some parameter $\varepsilon$.
\STATE \quad Output $\what^{est} = \what^{tv}\otimes \vone_{h \times h}$, where $\otimes$ denotes the Kronecker product.
\end{algorithmic}
\label{alg:sas}
\end{algorithm}

\subsection{Stage 1: Sorting}
The purpose of the sorting step is to rearrange the observed graph $G$ so that the rearranged empirical degrees are monotonically increasing. To this end, we compute the empirical degree
\begin{equation}
d_i \defequal \sum_{j=1}^n G_{ij},
\end{equation}
and define a permutation $\sigmahat$ such that $d_{\sigmahat(1)} < \ldots < d_{\sigmahat(n)}$. Then, we define a rearranged graph
\begin{equation}
\Ahat_{ij} = G_{\sigmahat(i)\sigmahat(j)},
\end{equation}
where an example is shown in \fref{fig:illustration}.

It is important to note that since the permutation $\sigmahat$ is defined by the empirical degrees, it could be different from the true permutation that defines the canonical graphon according to the node arrangement. To differentiate the empirical permutation $\sigmahat$ and the true permutation, we define $\sigma$ as the oracle permutation that sorts the node labels $(U_i)$ such that
$U_{\sigma(1)} < \ldots < U_{\sigma(n)}$. Correspondingly, we define the oracle ordered graph as
\begin{equation}
A_{ij} = G_{\sigma(i)\sigma(j)}.
\end{equation}

\subsection{Stage 2: Smoothing}
\textbf{Network Histogram Estimation}\\
Once the graph is rearranged to have monotonically increasing degrees, the graphon estimation problem becomes finding a smooth surface that best fits $(\Ahat_{ij})$. To this end, we consider a simplified version of the stochastic blockmodel approximation \cite{Airoldi_Costa_Chan_2013} which approximates the continuous graphon using a piecewise constant function. More precisely, the stochastic blockmodel approximation defines
\begin{align}
\Hhat_{ij} &= \frac{1}{h^2} \sum_{i_1=1}^h \sum_{j_1=1}^h \Ahat_{ih+i_1,jh+j_1}, \label{eq:Hhat def}
\end{align}
and correspondingly
\begin{align}
H_{ij}     &= \frac{1}{h^2} \sum_{i_1=1}^h \sum_{j_1=1}^h A_{ih+i_1,jh+j_1}, \label{eq:H def}
\end{align}
for some parameter $h > 0$ denoting the size of each block.

Equations \eref{eq:Hhat def} and \eref{eq:H def} indicate that the stochastic blockmodel approximations $(\Hhat_{ij})$ and $(H_{ij})$ are the histograms of $(\Ahat_{ij})$ and $(A_{ij})$, respectively. Since all function values in the same block are identical, the effective degrees of freedom in $(\Hhat_{ij})$ and $(H_{ij})$ are $k \times k$ instead of $n \times n$, where $k = \lfloor n/h \rfloor$ is the number of blocks.

\textbf{Total Variation Minimization}\\
While the network histogram estimation step is consistent, the decay rate of the error can be further improved by introducing a total variation minimization step.

The total variation minimization step is based on a sparsity assumption of the true graphon $w$. Analogous to natural images, we assume that graphons are \emph{sparse} in the gradients. Discretizing the continuous graphon $w$ into a $k \times k$ grid, the assumption suggests that $w$ needs to have a small total variation
\begin{equation}
\|w\|_{TV} = \sum_{i=1}^{k} \sum_{j=1}^{k} \sqrt{\left(\frac{\partial w}{\partial x}\right)_{ij}^2 + \left(\frac{\partial w}{\partial y}\right)_{ij}^2},
\end{equation}
where $\frac{\partial w}{\partial x}$ and $\frac{\partial w}{\partial y}$ denote the horizontal and vertical finite difference of $w$, respectively.

Using the total variation concept, the refinement step can be posed as the following minimization problem:
\begin{equation}
\what^{tv} = \argmin{\widehat{r}} \;\; \| \widehat{r} \|_{TV}, \quad \subjectto \| \widehat{r} - \Hhat \|_2 \le \varepsilon,
\label{eq:tv problem}
\end{equation}
where $\|\cdot\|_2$ is the matrix Frobenius norm, and $\varepsilon > 0$ is a parameter that controls the fidelity between the total variation solution $\widehat{r}$ and the histogram $\Hhat$.
%
%For completeness, we also define the problem corresponding to the oracle permutation:
%\begin{equation*}
%w^{tv} = \argmin{r} \;\; \| r \|_{TV}, \quad \subjectto \| r - H \|_2 \le \varepsilon.
%\end{equation*}
%
To solve the minimization problem \eref{eq:tv problem}, we use the alternating direction method of multipliers (ADMM) \cite{Chan_Khoshabeh_Gibson_2011}.

We remark that the size of $\what^{tv}$ is $k \times k$. To ensure that the final estimate have the same size as the true graphon, we define the final estimate as
\begin{equation}
\what^{est} = \what^{tv} \otimes \vone_{h \times h},
\end{equation}
where $\vone_{h \times h}$ denotes an all 1 matrix of size $h \times h$, and $\otimes$ denotes the Kronecker product operator. Therefore, the final estimate $\what_{est}$ has a size $n \times n$.

\subsection{Complexity}
The complexity of the SAS algorithm can be analyzed by considering each step individually. In computing the empirical degree distribution, $\calO(n)$ additions are used. The sorting procedure, in general, requires $\calO(n \log n)$ comparisons. Therefore, the complexity for sorting is about $\calO(n \log n)$ multiplications plus $\calO(n)$ additions. Next, for the histogram computation, computing each value of the bin requires $\calO(h^2)$ additions, and there are $k^2 = (n / h)^2$ bins. Thus a total of $\calO( n^2 )$ additions are needed. Finally, the total variation minimization is solved on a $k \times k$ array. Thus, the complexity of the ADMM step is $\calO(k^2 \log k^2)$. (See \cite{Chan_Khoshabeh_Gibson_2011} for discussions.) Combing these results we can show that the overall complexity of the SAS algorithm is $\calO(n \log n + k^2 \log k^2)$ multiplications plus $\calO(n^2)$ additions.

\section{Theoretical Properties of the SAS Algorithm}
In this section we discuss the statistical consistency of the proposed SAS algorithm.

Analyzing the consistency of the SAS algorithm is equivalent to determining an upper bound of the error
\begin{align}
&\MSE \,\defequal\, \frac{1}{n^2} \E\left[\| \what^{est} - w \|_2^2\right] \notag\\
&= \frac{1}{n^2} \Big( \E\left[ h^2 \| \what^{tv} - H^w \|_2^2 \right] + \E\left[\|H^w \otimes \vone_{h\times h} - w\|_2^2\right] \notag\\
&\quad + 2\E\left[(\what^{tv} - H^w)^T(H^w \otimes \vone_{h\times h} - w)\right] \Big), \label{eq:MSE}
\end{align}
where $H^w$ is the histogram approximation of $w$:
\begin{equation}
H^w_{ij} = \frac{1}{h^2} \sum_{i_1=1}^h \sum_{j_1=1}^h w_{ih+i_1,jh+j_1}.
\label{eq:Hw}
\end{equation}

Before we proceed, we note that the second expectation in \eref{eq:MSE} is a classical result of approximating a continuous function by step functions. The bound is given in the following Lemma.
\begin{lemma}[Piecewise Constant Function Approximation]
\label{lemma:step}
Let $w \in [0,1]^{n \times n}$ be the true graphon and let $H^w \in [0,1]^{k \times k}$ be the histogram approximation defined in \eref{eq:Hw}. Then,
\begin{equation}
\|H^w \otimes \vone_{h\times h} - w\|_2^2 \le \frac{C'}{k^2},
\end{equation}
where $C'$ is a constant independent of $n$.
\end{lemma}
Therefore, it remains to find an upper bound of $\| \what^{tv} - H^w \|_2^2$. (The last expectation in \eref{eq:MSE} can be bounded using Cauchy's inequality.) In the following subsections, we discuss how each step of the SAS algorithm contributes to this upper bound.

\subsection{Consistency of empirical degree sorting}
To establish the consistency of the empirical degree sorting, we must first establish the relationship between the oracle permutation $(\sigma(i))$ and the oracle degree $(d_{\sigma(i)})$.

\begin{lemma}
\label{thm:degree sorting}
Let $\sigma(i)$ be the oracle permutation such that $U_{\sigma(1)} < U_{\sigma(2)} < \ldots < U_{\sigma(n)}$. Let $g(u) = \int_0^1 w(u, v) dv$, and assume that there exists constants $L_1 > 0$ and $L_2 > 0$ such that
\begin{equation}
L_2 | x - y | \le | g(x) - g(y) | \le L_1 |x - y|,
\label{eq:lipschitz both side}
\end{equation}
for any $0 \le x \le 1$ and $0 \le y \le 1$. Then, the following result holds.

If $\left| \frac{\sigma(i)}{n} - \frac{\sigma(j)}{n} \right| < \frac{1}{6L_1}\sqrt{\frac{\log n}{n}}$, then
\begin{equation}
\label{eq:|di-dj| bound}
\left| d_{\sigma(i)} - d_{\sigma(j)} \right| < \sqrt{\frac{\log n}{n}},
\end{equation}
with probability at least $1-8e^{-\frac{1}{18L_1^2}\log n}$.

Conversely, if \eref{eq:|di-dj| bound} holds with probability at least $1-8e^{-\frac{1}{18L_1^2}\log n}$, then
\begin{equation}
\left| \frac{\sigma(i)}{n} - \frac{\sigma(j)}{n} \right| < \sqrt{\frac{\log n}{n}} \left[ \frac{1}{3L_1} + \frac{1}{3L_1L_2} + \frac{1}{L_2}\right],
\label{eq:|di-dj| bound converse}
\end{equation}
with probability at least $1-40e^{-\frac{1}{18L_1^2}\log n}$.
\end{lemma}

The interpretation of Lemma \ref{thm:degree sorting} is as follows. First, \eref{eq:lipschitz both side} is the two-sided Lipschitz condition, with Lipschitz constants $L_1$ and $L_2$. The Lipschitz condition enforces the degree distribution $g(u)$ to be well-behaved so that there is no abrupt transition for both $g$ and $g^{-1}$. Second, the forward statement suggests that if the oracle ordered indices have bounded differences, then correspondingly the empirical degrees should also have bounded differences. Conversely, \eref{eq:|di-dj| bound converse} suggests that if we can bound the difference in empirical degrees, then the difference in the true positions should also be bounded.

As an immediate consequence of Lemma \ref{thm:degree sorting}, we observe that for any fixed $i$, if we choose $j$ such that $\sigma(j) = \sigmahat(i)$, then the converse of Lemma \ref{thm:degree sorting} implies the following.
\begin{corollary}
If $\left| d_{\sigma(i)} - d_{\widehat{\sigma}(i)} \right| < \sqrt{\frac{\log n}{n}}$ holds with probability at least $1-8e^{-\frac{1}{18L_1^2}\log n}$, then
\begin{equation*}
\left| \frac{\sigma(i)}{n} - \frac{\widehat{\sigma}(i)}{n} \right| < C \sqrt{\frac{\log n}{n}}
\end{equation*}
holds with probability at least $1-40e^{-\frac{1}{18L_1^2}\log n}$, where $C = \frac{1}{3L_1} + \frac{1}{3L_1L_2} + \frac{1}{L_2}$ is a constant independent of $n$.
\end{corollary}
Therefore, if the error $\left| d_{\sigma(i)} - d_{\widehat{\sigma}(i)} \right|$ is small, then the error between $\sigma(i)$ and $\widehat{\sigma}(i)$ will also be small.

\subsection{Consistency of the histogram estimator}
During the histogram estimation step, the error associated with the empirical degree sorting is translated to the error between the empirical histogram $\Hhat$ and the ideal histogram $H$. This is reflected in the following lemma.
\begin{lemma}[Bounds on $\|\Hhat-H\|_{2}$]
\label{lemma:bound of H}
Let $w$ be the ground truth graphon and assume that $w$ is Lipschitz with constant $L > 0$. If $H$ and $\Hhat$ are defined according to \eref{eq:H def} and \eref{eq:Hhat def}, respectively, then
\begin{align}
\E[ \|\Hhat - H\|_{2}^2 ]  &\le \frac{k^4}{n^2} \left(2+4C^2 L^2 \frac{\log n}{n}\right) \notag \\
&\quad + k^2\left(4C^2 L^2 {\frac{\log n}{n}} \right),
\label{eq:bound of H}
\end{align}
where $C$ is a constant independent of $n$.
\end{lemma}
We also establish the relationship between $H$ and the step approximation $H^w$.
\begin{lemma}[Bounds on $\|H-H^w\|_{2}$]
\label{lemma:bound of H wstep}
Let $H^w$ be the step function approximation of the graphon $w$ and let $H$ be the histogram defined as \eref{eq:H def}. Then,
\begin{equation}
\E[ \|H - H^w \|_{2}^2] \le \frac{k^4}{n^2}.
\label{eq:bound of H wstep}
\end{equation}
\end{lemma}

\subsection{Consistency of total variation smoothing}
To analyze the total variation minimization step, we first observe that
\begin{equation}
\Hhat = H^w + \underset{\eta}{\underbrace{\Hhat - H}} + \underset{\rho}{\underbrace{H - H^w}} \label{eq:Hhat noise model}.
\end{equation}
Therefore, if we consider $H^w$ as the desired function to be estimated, and consider $\eta$ and $\rho$ as perturbations added to $H^w$, then $\Hhat$ can be regarded as a noisy observation of $H^w$. Consequently, by applying total variation minimization to \eref{eq:Hhat noise model}, we find a solution $\what^{tv}$ that best fits \eref{eq:Hhat noise model} and has the minimum total variation.

To characterize the solution of the total variation minimization problem, we first define the $s$-sparsity of the gradient of a function $H^w$.
\begin{definition}
A function $H^w \in [0,1]^{k \times k}$ is $s$-sparse in gradient if its gradient $\nabla H^w$ has at most $s$ non-zero entries.
\end{definition}
With this definition, we apply the following result in compressed sensing.
\begin{lemma}[\cite{Needell_Ward_2013} Theorem A]
\label{lemma:tv solution}
If $\Hhat = H^w + \eta + \rho$ with $\varepsilon^2 = \E[\|\eta + \rho\|_{2}^2]$, then the solution $\what^{tv}$ of
\begin{align*}
\what^{tv} = \argmin{\widehat{r}} \|\widehat{r}\|_{TV} \quad \subjectto \quad \| \widehat{r} - \Hhat\|_{2} \le \varepsilon,
\end{align*}
satisfies the condition
\begin{align*}
\| \what^{tv} - H^w \|_{2} \le \frac{\| \nabla H^w - (\nabla H^w)_s \|_{1}}{\sqrt{s}} + \varepsilon,
\end{align*}
where $(\cdot)_s$ denotes the function reconstructed from the $s$ most significant non-zero entries of the argument.
\end{lemma}

Lemma~\ref{lemma:tv solution} indicates that the error $\| \what^{tv} - H^w \|_{2}$ is controlled by the perturbation $\varepsilon$ and the sparse approximation error $\| \nabla H^w - (\nabla H^w)_s \|_{1}$. Since $\varepsilon^2 = \E[\|\eta + \rho\|_{2}^2]$, and $\eta$ and $\rho$ are defined according to \eref{eq:Hhat noise model}, $\varepsilon$ is upper bounded by \eref{eq:bound of H} and \eref{eq:bound of H wstep}. For the sparse approximation error term, in general $\| \nabla H^w - (\nabla H^w)_s \|_{1} \not= 0$ because $H^w$ is not necessarily $s$-sparse in gradient. However, in practice, many real world networks are sparse (\emph{i.e.} number of edges are much fewer than number of nodes). Therefore, for practical consideration it is often reasonable to assume that $H^w$ is $s$-sparse in gradient and so $\|\nabla H^w - (\nabla H^w)_s\|_1=0$.

\subsection{Overall consistency}
In summary, the overall consistency is given by the following theorem.

\begin{theorem}[Consistency of SAS algorithm]
\label{thm:consistency}
Let $w$ be the true graphon with the following properties: (i) $w$ is Lipschitz with constant $L>0$; (ii) $g(u) = \int_{0}^1 w(u,v)dv$ is Lipschitz as defined in Lemma~\ref{thm:degree sorting}; (iii) $H^w$ is $s$-sparse in gradient. Then, the MSE of the SAS estimator satisfies
\begin{equation}
\MSE \le \calO \left( \frac{\log n}{n} \right),
\end{equation}
and hence $\MSE \rightarrow 0$ as $n \rightarrow \infty$ and $k/n \rightarrow 0$, where $k$ is the number of blocks defined in \eref{eq:H def}.
\end{theorem}

\section{Experimental results}
After establishing the theoretical results, we now present simulation results of the proposed SAS algorithm.

\subsection{Simulations}
The first experiment considers a number of graphons listed in Table~\ref{table:list graphons}. The choices of these graphons are made to include both low rank and high rank graphons, where the rank is measured numerically using a $1000 \times 1000$ discretization of the continuous graphons. Among the 10 graphons listed in Table~\ref{table:list graphons}, we note that graphon no. 1 $w(u,v) = uv$ is a special case of the eigenmodel \cite{Hoff_2008}, graphon no. 5 $w(u,v) = 1/(1+\exp\{-10(u^2+v^2)\})$ is a variation of the logistic model presented in \cite{Chatterjee_2012}, and graphon no. 6 $w(u,v) = |u-v|$ is the latent distance model \cite{Hoff_Raftery_Handcock_2002}. Other graphons are chosen to demonstrate the robustness of the SAS algorithm.

\begin{table}[h]
\begin{tabular}{|c|c|c|}
\hline
ID   & $w(u,v)$ & $rank(w)$\\
\hline\hline
1    & $uv$                                                 & 1 \\
2    & $\exp\{-(u^{0.7} + v^{0.7})\}$                       & 1 \\
3    & $\frac{1}{4}\left[u^2+v^2+u^{1/2}+v^{1/2}\right]$    & 2 \\
4    & $\frac{1}{2}(u+v)$                                   & 2 \\
5    & $\frac{1}{1+\exp\{-10(u^2+v^2)\}}$                     & 10 \\
6    & $|u-v|$                                                & 1000 \\
7    & $\frac{1}{1+\exp\{-(\max(u,v)^2 + \min(u,v)^4)\}}$     & 1000 \\
8    & $\exp\{-\max(u,v)^{3/4}\}$                             & 1000 \\
9    & $\exp\{-\frac{1}{2}\left(\min(u,v) + u^{1/2} + v^{1/2}\right)\}$ & 1000 \\
10   & $\log(1+ 0.5\max(u,v))$                              & 1000\\
\hline
\end{tabular}
\caption{List of graphons for testing. The rank of $w$ is estimated from a $1000 \times 1000$ discretization of the graphon.}
\label{table:list graphons}
\end{table}

We compare the SAS algorithm with the universal singular value thresholding (USVT) algorithm \cite{Chatterjee_2012} and the stochastic blockmodel approximation algorithm \cite{Airoldi_Costa_Chan_2013}. These two algorithms are the existing methods that have provable consistency and are numerically efficient. However, since both of these two methods do not have a sorting step, we apply the sorting step of the SAS algorithm prior to running the two algorithms. For the choice of binwidth $h$, we set $h = \log n$ for the SAS algorithm, and an oracle $h$ that minimizes the MSE for the SBA algorithm (\emph{i.e.}, using the ground truth).

\begin{table}[t]
\centering
\setlength{\extrarowheight}{1.5pt}
{
\begin{tabular}{|c|ccc|}
\hline
         & \multicolumn{3}{|c|}{$n = 200$}   \\
\hline
ID       &   SAS (Proposed)  &  USVT \cite{Chatterjee_2012}  & SBA \cite{Airoldi_Costa_Chan_2013}  \\
\hline
1	 & 	 6.59e-04  $\pm$   5.18e-05 	 & 	 1.90e-03 $\pm$ 1.88e-04 	 & 	2.77e-03	$\pm$	1.60e-04	\\
2	 & 	 4.92e-04  $\pm$   6.81e-05 	 & 	 2.18e-03 $\pm$ 1.95e-04 	 & 	2.36e-03	$\pm$	1.97e-04	\\
3	 & 	 6.95e-04  $\pm$   7.52e-05 	 & 	 3.12e-03 $\pm$ 2.32e-04 	 & 	5.08e-03	$\pm$	2.26e-04	\\
4	 & 	 6.48e-04  $\pm$   5.30e-05 	 & 	 3.51e-03 $\pm$ 1.93e-04 	 & 	2.77e-03	$\pm$	1.49e-04	\\
5	 & 	 9.74e-05  $\pm$   2.76e-05 	 & 	 3.15e-03 $\pm$ 8.76e-19 	 & 	3.13e-03	$\pm$	3.31e-04	\\
6	 & 	 4.29e-02  $\pm$   9.27e-05 	 & 	 8.91e-02 $\pm$ 1.23e-03 	 & 	4.37e-02	$\pm$	1.20e-04	\\
7	 & 	 4.81e-04  $\pm$   7.50e-05 	 & 	 2.40e-03 $\pm$ 1.77e-04 	 & 	2.71e-03	$\pm$	2.09e-04	\\
8	 & 	 9.38e-04  $\pm$   1.21e-04 	 & 	 6.27e-03 $\pm$ 1.58e-03 	 & 	1.52e-03	$\pm$	1.52e-04	\\
9	 & 	 6.50e-04  $\pm$   7.73e-05 	 & 	 2.87e-03 $\pm$ 2.32e-04 	 & 	3.96e-03	$\pm$	3.25e-04	\\
10	 & 	 7.67e-04  $\pm$   1.01e-04 	 & 	 4.74e-03 $\pm$ 6.25e-04 	 & 	1.13e-03	$\pm$	1.23e-04	\\
\hline
Average  &   4.83e-03  $\pm$   7.43e-05		 &   1.19e-02 $\pm$ 4.65e-04	 &   6.91e-03 $\pm$ 1.99e-04\\
\hline
\hline
& \multicolumn{3}{|c|}{$n=1000$}  \\
\hline
ID       &   SAS (Proposed)  &  USVT \cite{Chatterjee_2012}  & SBA \cite{Airoldi_Costa_Chan_2013} \\
\hline
1	&	            8.56e-05   $\pm$   3.42e-06 	 & 	 3.86e-04$\pm$1.70e-05 	 & 	9.00e-04	$\pm$	1.70e-05	 \\
2	&	            7.12e-05   $\pm$   5.92e-06 	 & 	 4.46e-04$\pm$1.84e-05 	 & 	1.39e-03	$\pm$	3.99e-05	 \\
3	&	            9.60e-05   $\pm$   5.78e-06 	 & 	 9.69e-04$\pm$2.67e-05 	 & 	8.66e-04	$\pm$	1.90e-05	 \\
4	&	            7.82e-05   $\pm$   5.17e-06 	 & 	 8.83e-04$\pm$2.47e-05 	 & 	1.43e-03	$\pm$	2.63e-05	 \\
5	&	            1.09e-05   $\pm$   1.66e-06 	 & 	 8.69e-05$\pm$7.03e-06 	 & 	1.60e-03	$\pm$	3.45e-05	 \\
6	&	            4.19e-02   $\pm$   9.58e-06 	 & 	 8.42e-02$\pm$1.70e-04 	 & 	4.22e-02	$\pm$	1.42e-05	 \\
7	&	            8.48e-05   $\pm$   7.47e-06 	 & 	 6.76e-04$\pm$1.81e-05 	 & 	1.21e-03	$\pm$	3.65e-05	 \\
8	&	            1.73e-04   $\pm$   1.30e-05 	 & 	 1.66e-03$\pm$4.56e-05 	 & 	6.81e-04	$\pm$	2.14e-05	 \\
9	&	            1.02e-04   $\pm$   5.15e-06 	 & 	 1.26e-03$\pm$3.01e-05 	 & 	1.15e-03	$\pm$	3.44e-05	 \\
10	&	            1.37e-04   $\pm$   1.02e-05 	 & 	 1.24e-03$\pm$3.30e-05 	 & 	7.38e-04	$\pm$	1.67e-05	 \\
\hline
Average  &       	0.27e-03	$\pm$   6.74e-06	 &	 9.18e-03$\pm$3.91e-05	 &   5.22e-03$\pm$2.06e-05 \\
\hline
\end{tabular}
}
\caption{Mean squared error (average $\pm$ std. dev.) comparisons between theSAS algorithm, the USVT algorithm \cite{Chatterjee_2012}, and the SBA algorithm \cite{Airoldi_Costa_Chan_2013}. MSEs are averaged over 50 independent trials.}
\label{table:mse}
\end{table}

The results of the experiment are shown in Table~\ref{table:mse}, where we report the mean squared error (MSE) of the estimated graphons using the SAS algorithm, the USVT algorithm and the SBA algorithm. To reduce the random fluctuations caused by independent realizations of the random graphs, we average the MSE over 50 independent trials. Two cases of graph sizes are considered: $n = 200$ and $n = 1000$. The results show that the SAS algorithm in general outperforms the USVT algorithm and the SBA algorithm. Averaged over the 10 testing graphons, we see that the SAS algorithm achieves the lowest MSE among all three methods.

\begin{figure}[t]
\centering
\begin{tabular}{ccc}
SAS (Proposed) & USVT & SBA\\
\includegraphics[width=0.28\linewidth]{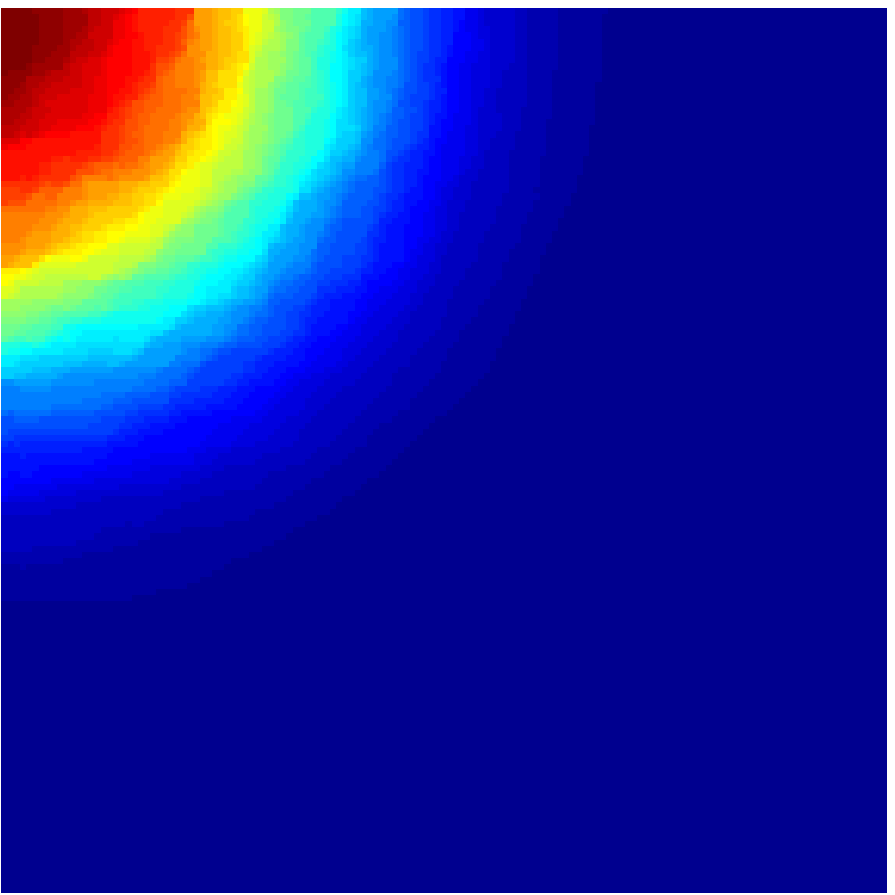}&
\includegraphics[width=0.28\linewidth]{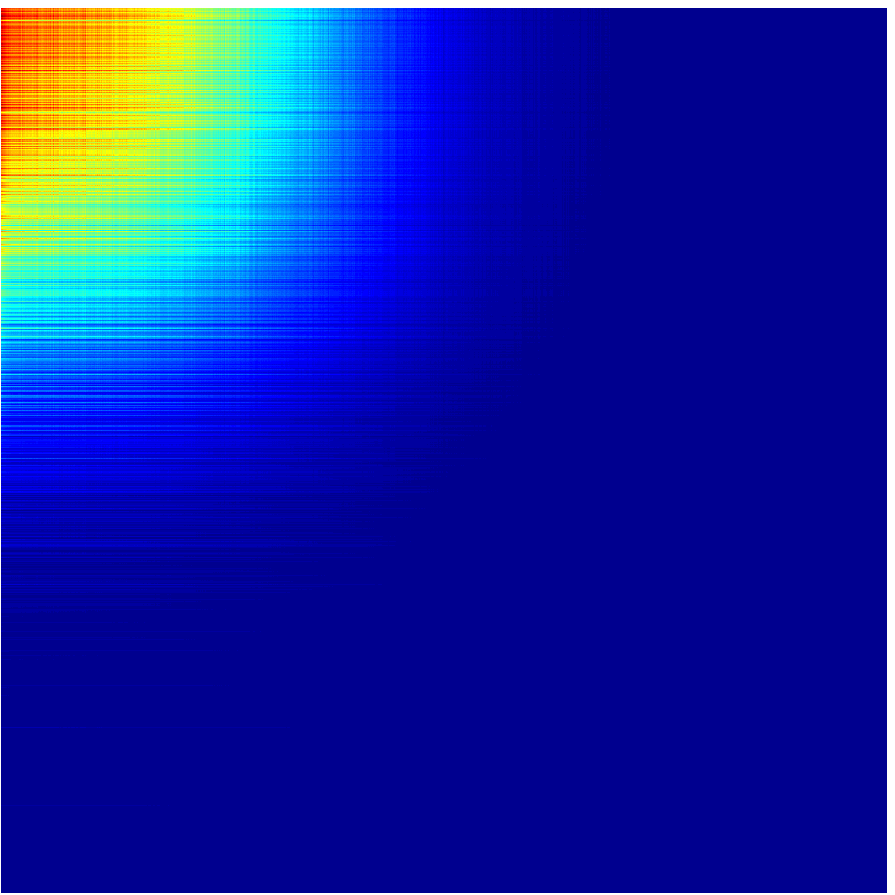}&
\includegraphics[width=0.28\linewidth]{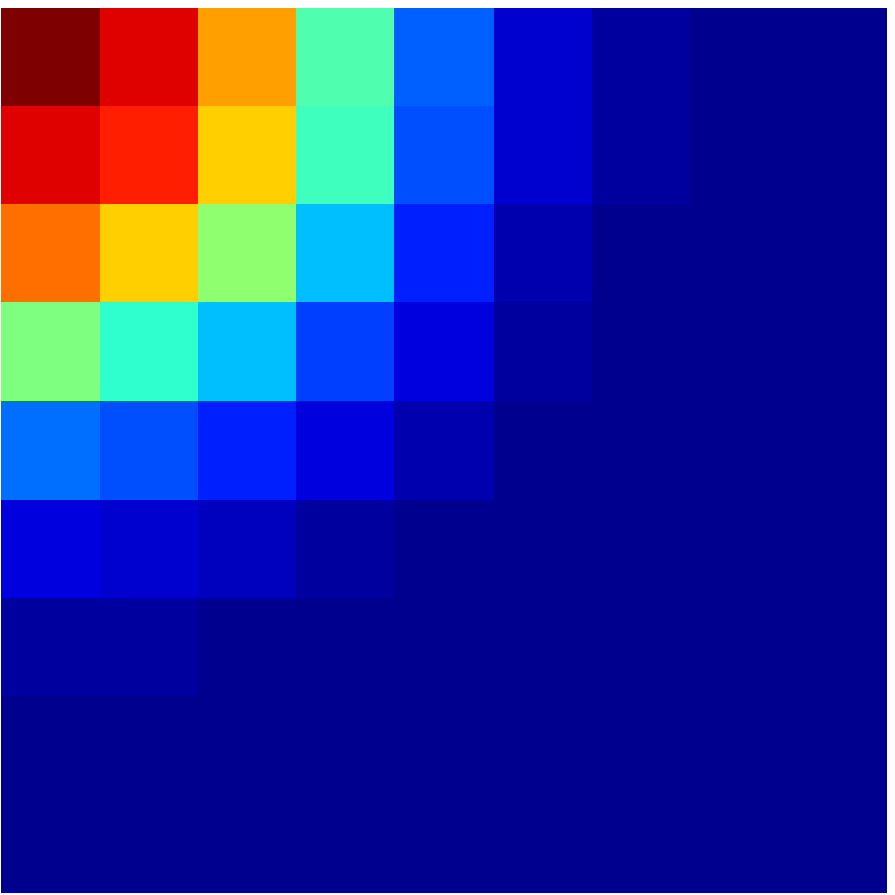}\\
(a) $1.09\times10^{-5}$ & (b) $8.69\times10^{-5}$ & (c) $1.60\times10^{-3}$\\
\includegraphics[width=0.28\linewidth]{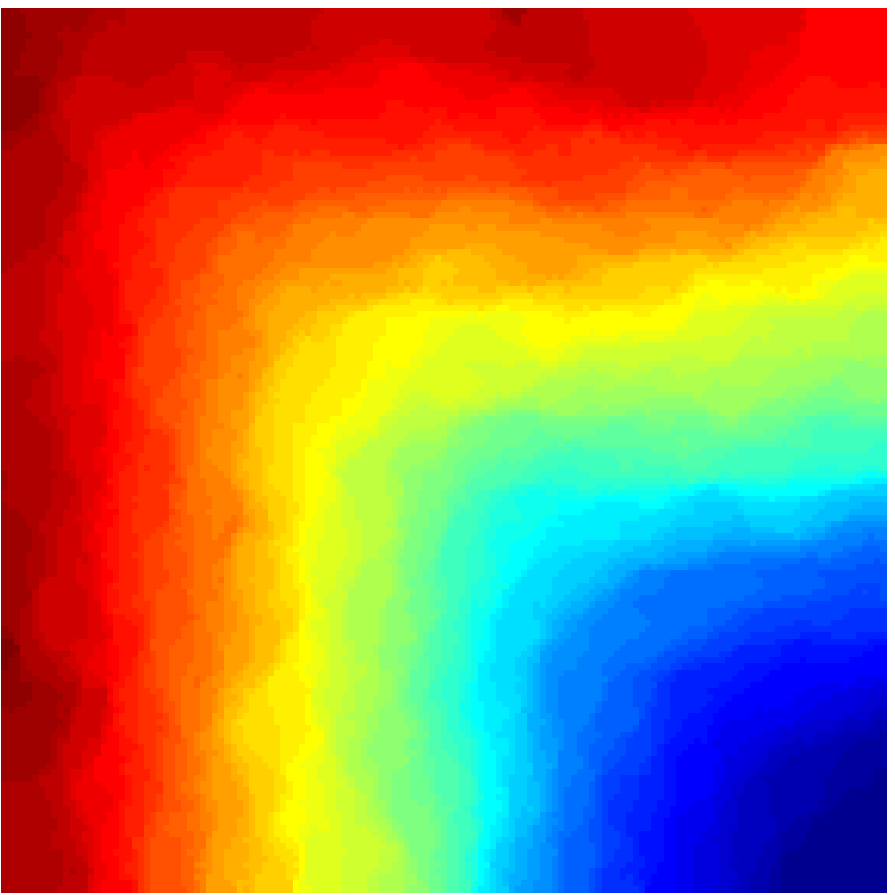}&
\includegraphics[width=0.28\linewidth]{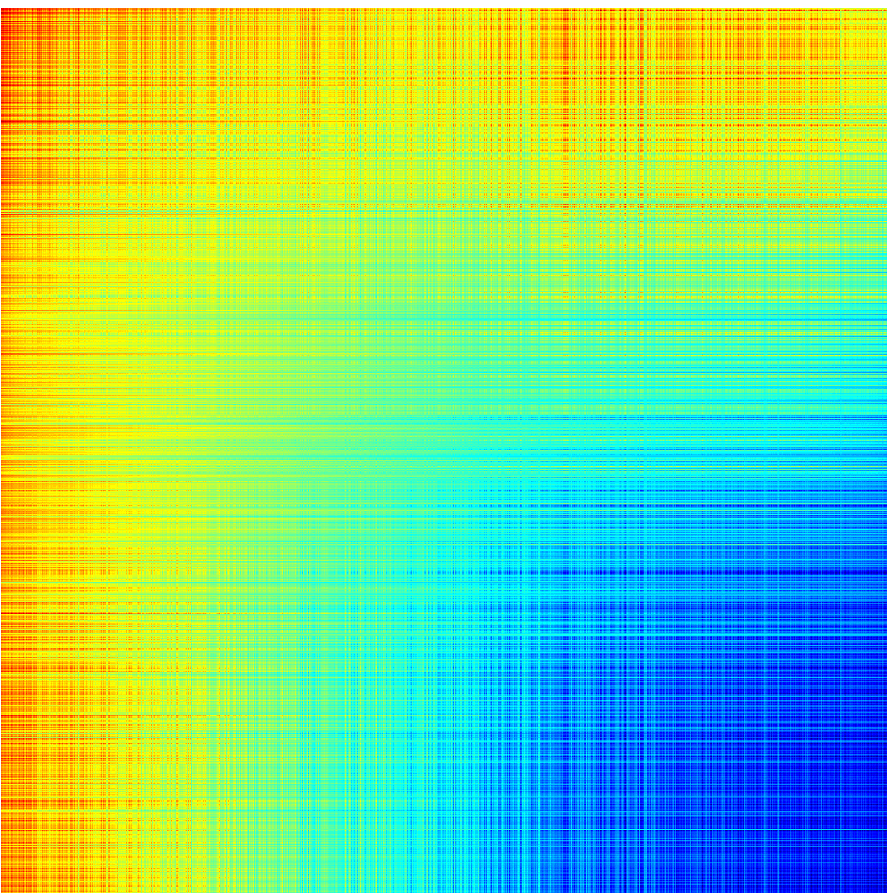}&
\includegraphics[width=0.28\linewidth]{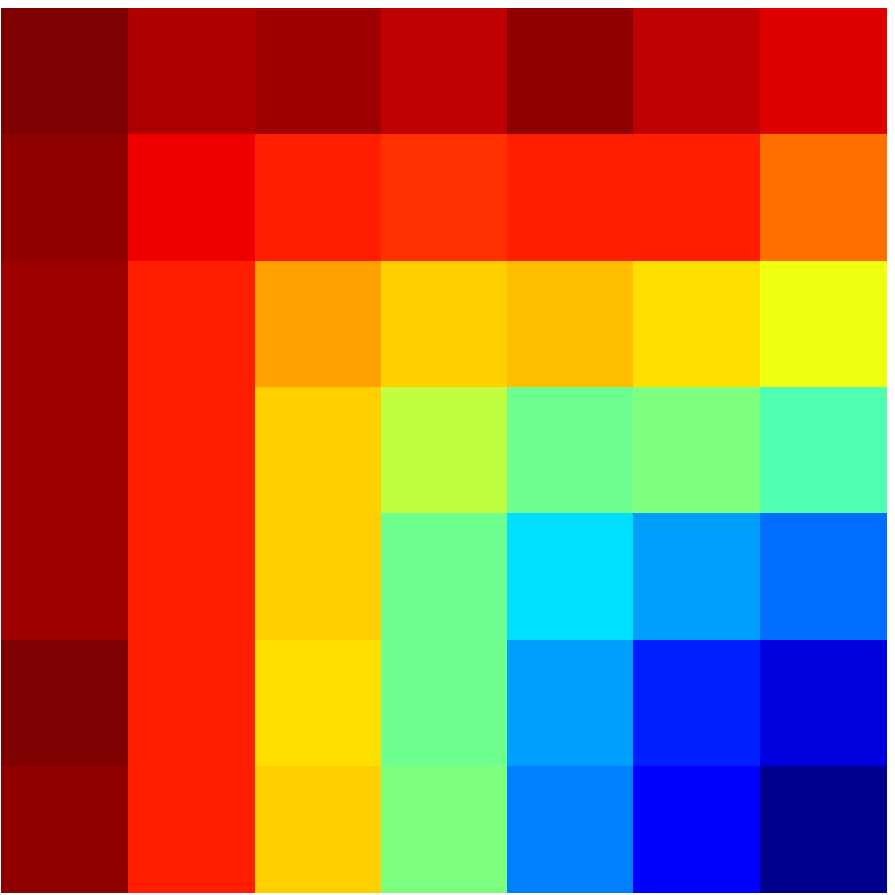}\\
(d) $1.37\times10^{-4}$ & (e) $1.24\times10^{-3}$ & (f) $7.38\times10^{-4}$\\
\end{tabular}
\caption{Comparisons between the SAS algorithm, the USVT algorithm \cite{Chatterjee_2012}, and the SBA algorithm \cite{Airoldi_Costa_Chan_2013}. Numbers indicate the mean squared error. (a)-(c): Graphon 5; (d)-(f): Graphon 10. SAS algorithm uses $h = \log n$. SBA algorithm uses an oracle $h$ that minimizes the MSE. In this example, we set $n = 1000$.}
\label{fig:results}
\end{figure}

\fref{fig:results} displays two examples of the estimated graphons. As shown in the figure, we see that while the USVT algorithm returns
a reasonable estimate for graphon no.5 (which has a low rank), it returns a relatively worse estimate for graphon no. 10 (which has a high rank). Looking at the SBA algorithm, it is evident that using the oracle binwidth $h$, the average MSE is lower than that of USVT. However, the SBA algorithm tends to return a graphon with few communities. This is not favorable if the network has non-block structures. In contrast, the SAS algorithm returns results with lower MSE, and retains important features of the true graphons.

\begin{figure}[t]
\centering
\includegraphics[width=0.75\linewidth]{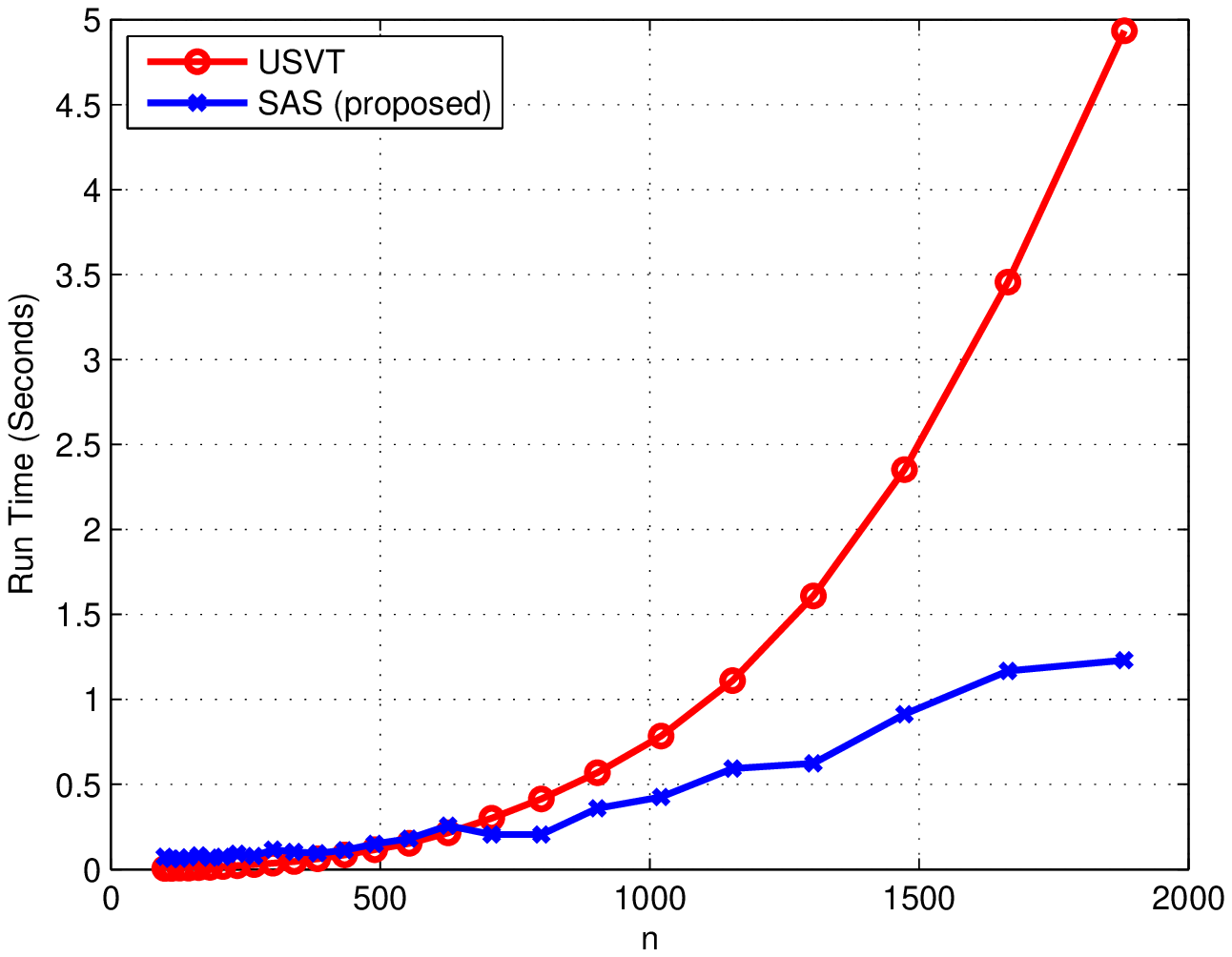}
\caption{Run time comparison between USVT \cite{Chatterjee_2012} and the SAS algorithm (averaged over 10 graphons shown in Table~\ref{table:list graphons}).}
\label{fig:runtime}
\end{figure}

In \fref{fig:runtime} we show the runtime comparison between the SAS algorithm and the USVT algorithm. Both algorithms are implemented on an Intel 3.5GHz machine with 16GB RAM, Windows 7 / MATLAB R7.12.0 platform. The runtime plot indicates that the SAS algorithm has a significantly lower complexity than the USVT algorithm.

\subsection{Real data analysis}
As an application of the proposed SAS algorithm, we consider the problem of estimating graphons from real-world networks. For this purpose, we consider the collaboration network of arXiv astro physics (ca-AstroPh) and the who-trusts-whom network of Epinions.com (soc-Epinions1) from Stanford Large Network Dataset Collection\footnote{http://www.cise.ufl.edu/research/sparse/matrices/SNAP/}. The ca-AstroPh network is a symmetric binary graph consisting of $1.8\times10^4$ nodes and $3.9\times10^5$ edges, whereas the soc-Epinions-1 network is an unsymmetrical binary graph consisting of $7.5\times10^4$ nodes and $5.1\times10^5$ edges. For both networks, we randomly permute the rows and columns to simulate the raw data scenario where nodes are initially unordered.

\begin{figure}[!]
\centering
\begin{tabular}{cc}
\includegraphics[width=0.43\linewidth]{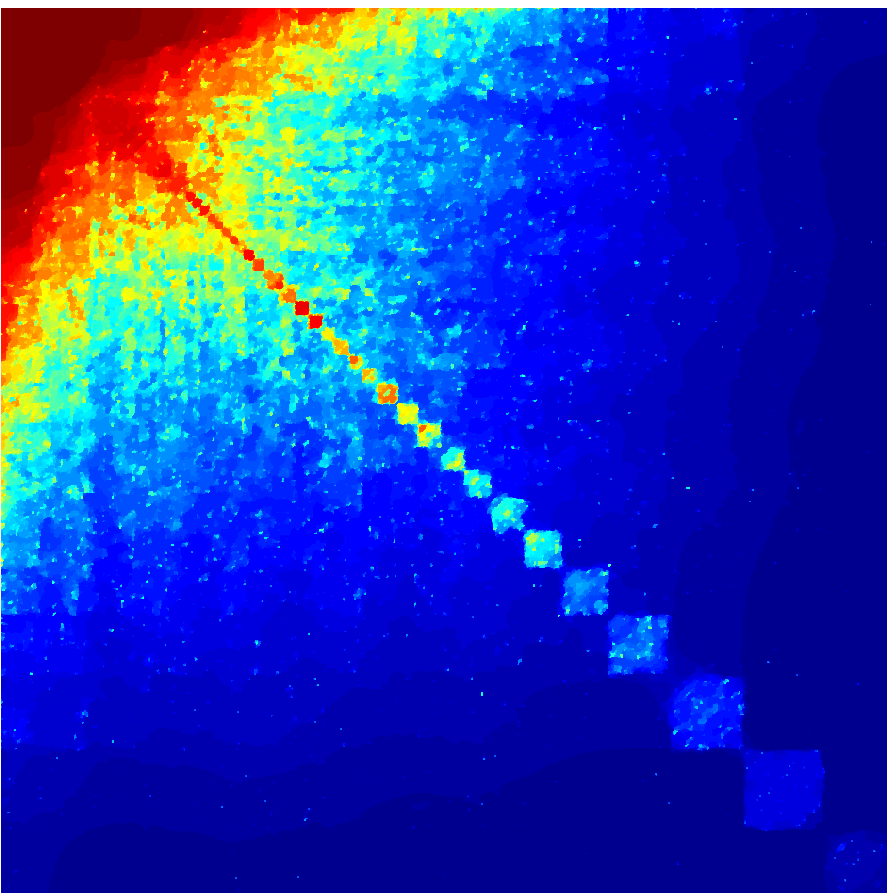}&
\includegraphics[width=0.43\linewidth]{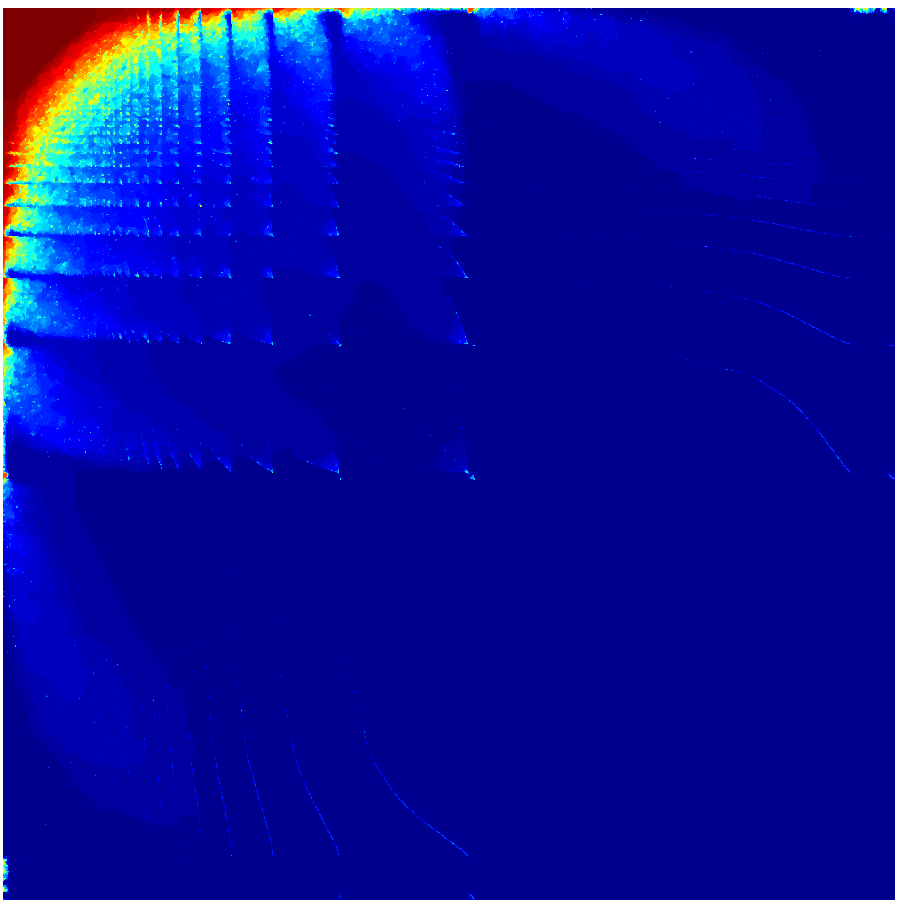}
\end{tabular}
\caption{Estimated graphons for real networks. Left: Collaboration network of arXiv astro physics (ca-AstroPh) $n = 1.8\times10^4$. Right: who-trusts-whom network of Epinions.com (soc-Epinions1) $n = 7.5\times10^4$.}
\label{fig:results2}
\end{figure}

\fref{fig:results2} shows the results of the SAS algorithm. For the ca-AstroPh network, the graphon shows close collaborations among a group of people concentrated around the top left corner of the graphon. It also shows a number of small communities along the diagonal. For the soc-Epinions1 network, the graphon indicates that there are some influential nodes which consistently interact among themselves. These can be seen from the repeated patterns of the graphon.

We remark that or the ca-AstroPh network ($n = 1.8\times10^4$) and the soc-Epinions-1 network ($n = 7.5\times10^4$), the estimations are completed in 20 seconds and 170 seconds, respectively, on a PC using an unoptimized MATLAB code. This provides a strong indication of the scalability of the SAS algorithm to larger networks.

\section{Conclusion}
The Sorting-And-Smoothing (SAS) algorithm is a consistent and efficient graphon estimation algorithm. The SAS algorithm consists of two steps. In the first step, the observed graph is rearranged so that the degrees are monotonically increasing. In the second step, a histogram estimation and a total variation minimization is applied to estimate a smooth surface that best fits the observed data. The SAS algorithm is evaluated on both simulation data and real network data. Our simulation results indicate that the SAS algorithm outperforms the universal singular value thresholding algorithm and the stochastic blockmodel approximation algorithm. On large-scale real networks, the SAS algorithm returns consistent graphon estimates.

\textbf{Code}. Available at: https://github.com/airoldilab/SAS

\textbf{Acknowledgments}. The authors thank J. J. Yang and C. Q. Han for useful discussions. SHC is partially supported by a Croucher Foundation Postdoctoral Research Fellowship. EMA is partially supported by NSF CAREER award IIS-1149662, AROMURI award W911NF-11-1-0036, and an Alfred P. Sloan Research Fellowship.

\appendix
\section{Total Variation Minimization}
The purpose of this appendix is to provide a brief summary of the ADMM algorithm used to solve the total variation minimization problem. For detailed discussions, we refer the readers to \cite{Chan_Khoshabeh_Gibson_2011}.

The problem of interest is the following minimization problem:
\begin{equation}
\what^{tv} = \argmin{\rhat} \;\; \|\rhat\|_{TV}, \;\;\subjectto\;\; \|\rhat - \Hhat\|_2^2 \le \varepsilon.
\label{eq:admm,problem0}
\end{equation}
To solve this minimization problem, we consider an equivalent unconstrained problem
\begin{equation}
\what^{tv} = \argmin{\rhat} \;\; \frac{\mu}{2}\|\rhat - \Hhat\|_2^2 + \|\rhat\|_{TV},
\label{eq:admm,problem}
\end{equation}
for some parameter $\mu$. It can be shown that for any fixed $\varepsilon$, there exists $\mu$ such that the solutions of \eref{eq:admm,problem0} and \eref{eq:admm,problem} coincides. Thus, it suffices to solve \eref{eq:admm,problem}.

In \eref{eq:admm,problem}, the total variation norm is defined as
\begin{equation*}
\|\rhat\|_{TV} = \| D \, \rhat \|_{*},
\end{equation*}
where $D: \R^{k \times k} \rightarrow \R^{2k \times k}$ consists of the horizontal and vertical finite difference operators:
\[D\rhat = \begin{bmatrix} D_x \rhat \\ D_y \rhat \end{bmatrix},\]
where
\begin{align*}
D_x \rhat &=
\begin{cases} \rhat_{i+1,j} - \rhat_{i,j}, &\quad i = 1,\ldots,k-1,\\
\rhat_{1,j} - \rhat_{k,j} &\quad i = k,
\end{cases}\\
D_y \rhat &=
\begin{cases} \rhat_{i,j+1} - \rhat_{i,j}, &\quad j = 1,\ldots,k-1,\\
\rhat_{i,1} - \rhat_{i,k} &\quad j = k,
\end{cases}
\end{align*}
and $\|\cdot\|_{*}$ is the norm defined as
\begin{equation*}
\left\| \begin{bmatrix} z_1 \\ z_2 \end{bmatrix} \right\|_{*} = \sum_{i=1}^k \sum_{j=1}^k \sqrt{(z_1)_{ij}^2 + (z_2)_{ij}^2}.
\end{equation*}
Using the definition of the total variation norm, \eref{eq:admm,problem} can be written as
\begin{equation}
\what^{tv} = \argmin{\rhat} \;\; \frac{\mu}{2}\|\rhat - \Hhat\|_2^2 + \|D \rhat\|_{*}.
\label{eq:admm,problem2}
\end{equation}

The difficulty in solving \eref{eq:admm,problem2} is that the quadratic term $\|\rhat - \Hhat\|_2^2$ is differentiable whereas the total variation term $\|D \rhat\|_{*}$ is not differentiable. In order to split the two terms, we introduce an auxiliary variable $\uhat$ and consider an equivalent constrained problem
\begin{equation}
\what^{tv} = \argmin{\rhat,\uhat} \;\; \frac{\mu}{2}\|\rhat - \Hhat\|_2^2 + \|\uhat\|_{*}, \quad \subjectto\quad \uhat = D\rhat.
\label{eq:admm,problem3}
\end{equation}
Since the constraint $\uhat = D\rhat$ must be satisfied at the optimal point, the solution of \eref{eq:admm,problem3} is the same as the solution of \eref{eq:admm,problem2}.

The idea of the ADMM algorithm is to consider the augmented Lagrangian function of \eref{eq:admm,problem3}, which is defined as
\begin{equation*}
\calL(\rhat,\uhat,\zhat) = \argmin{\rhat,\uhat} \;\; \frac{\mu}{2}\|\rhat - \Hhat\|_2^2 + \|\uhat\|_{*} - \zhat^T(\uhat - D\rhat) + \frac{\rho}{2}\|\uhat - D\rhat\|_2^2,
\end{equation*}
where $\zhat \in \R^{2k \times 1}$ is the Lagrange multiplier associated with the constraint $\uhat = D\rhat$, and $\frac{\rho}{2}\|\uhat - D\rhat\|_2^2$ is a quadratic penalty.

An important fact of the ADMM algorithm is that the optimum point of \eref{eq:admm,problem3} is also the saddle point of $\calL$, which can be determined iteratively by solving the following subproblems, with the $k$-th iteration as
\begin{align}
\rhat_{k+1} &= \argmin{\rhat} \;\; \calL(\rhat,\uhat_k,\zhat_k) = (\mu + \rho D^T D)^{-1}(\mu H + \rho D^T \uhat_k - D^T \zhat_k), \label{eq:ADMM,rsub}\\
\uhat_{k+1} &= \argmin{\uhat} \;\; \calL(\rhat_{k+1},\uhat,\zhat_k) = \max\left\{ \vhat_k - \frac{1}{\rho}, 0\right\} \cdot \begin{bmatrix}\vhat_k^{(1)} \\ \vhat_k^{(2)}\end{bmatrix} \cdot \slash \vhat_k,
\label{eq:ADMM,usub}\\
\zhat_{k+1} &= \argmax{\zhat} \;\; \calL(\rhat_{k+1},\uhat_{k+1},\zhat) = \zhat_k - \rho(\uhat_{k+1} - D\rhat_{k+1}),
\label{eq:ADMM,zsub}
\end{align}
where
$$
\begin{bmatrix} \vhat_k^{(1)} \\ \vhat_k^{(2)} \end{bmatrix} = D\rhat_{k} + \frac{1}{\rho} \zhat_{k}
 \quad\mbox{and}\quad
\vhat_k = \sqrt{ (\vhat_k^{(1)})^2 + (\vhat_k^{(2)})^2}.
$$

The minimization problems \eref{eq:ADMM,rsub}, \eref{eq:ADMM,usub} and \eref{eq:ADMM,zsub} are known as the $\rhat$-, $\uhat$-, and $\zhat$-subproblems, respectively. To solve the $\rhat$-subproblem, we note that the operator $D$ is a block-circulant matrix. Therefore, the Fourier transform matrix $F$ can be used to diagonalize $D^TD$ as $D^TD = F S F^H$, where $S$ is the eigenvalue matrix. Consequently, the inverse $(\mu + \rho D^T D)^{-1}$ can be efficiently executed using the fast Fourier transform operations. The $\uhat$-subproblem $\argmin\, \calL(\rhat_{k+1},\cdot,\zhat_k)$ involves solving a sum of separable single-variable problems of which the solution is given by \eref{eq:ADMM,usub}. \eref{eq:ADMM,usub} is also known as the shrinkage solution, and the operations ``$\slash$'' and ``$\cdot$'' are elementwise operations. The $\zhat$-subproblem can be interpreted as the steepest ascent of $\calL$ along the $\zhat$ direction.

The overall complexity of the ADMM algorithm is upper bounded by the $\rhat$-subproblem where a 2-dimensional fast Fourier transform is involved. Since the fast Fourier transform has a complexity of $k^2 \log k^2$, the complexity of the ADMM algorithm is $\calO(k^2 \log k^2)$.

\section{Proofs}
\subsection{Proof of Lemma 1}
We first prove the forward direction. Suppose that $\left| \frac{\sigma(i)}{n} - \frac{\sigma(j)}{n} \right| < \delta$ for some $\delta > 0$. Then,
\begin{align*}
&\Pb\left( \left| U_{\sigma(i)} - U_{\sigma(j)} \right| > 3 \delta \right)\\
&\le \Pb\left( \left| U_{\sigma(i)} - \frac{\sigma(i)}{n} \right| + \left| U_{\sigma(j)} - \frac{\sigma(j)}{n} \right| + \left|\frac{\sigma(i)}{n} - \frac{\sigma(j)}{n}\right|> 3 \delta \right)\\
&\le \Pb\left( \left| U_{\sigma(i)} - \frac{\sigma(i)}{n} \right| + \left| U_{\sigma(j)} - \frac{\sigma(j)}{n} \right| > 2 \delta \right) \\
&\le \Pb\left( \left| U_{\sigma(i)} - \frac{\sigma(i)}{n} \right| > \delta \right) + \Pb\left( \left| U_{\sigma(j)} - \frac{\sigma(j)}{n} \right| > \delta \right)\\
&\overset{(a)}{\le} 2\exp\{-2n\delta^2\} + 2\exp\{-2n\delta^2\}\\
&= 4\exp\{-2n\delta^2\},
\end{align*}
where $(a)$ is due to Dvoretzky. Consequently,
\begin{align*}
\Pb\left( \left| g(U_{\sigma(i)}) - g(U_{\sigma(j)}) \right| > 3 L_1 \delta \right)
&\overset{(b)}{\le} \Pb\left( \left| U_{\sigma(i)} - U_{\sigma(j)} \right| > 3 \delta \right)
\le 4\exp\{-2n\delta^2\},
\end{align*}
where $(b)$ is due to Lipschitz. Therefore,
\begin{align*}
&\Pb\left( \left| d_{\sigma(i)} - d_{\sigma(j)} \right| > 6 L_1 \delta \;\Big|\; U_{\sigma(i)}, U_{\sigma(j)}\right) \\
&\le \Pb\Big( \left| d_{\sigma(i)} - g(U_{\sigma(i)}) \right| + \left| d_{\sigma(j)} - g(U_{\sigma(j)})\right| \\
&\quad + \left| g(U_{\sigma(i)}) - g(U_{\sigma(j)}) \right|> 6 L_1 \delta \;\Big|\; U_{\sigma(i)}, U_{\sigma(j)}\Big)\\
&\overset{(c)}{\le_p} \Pb\left( \left| d_{\sigma(i)} - g(U_{\sigma(i)}) \right| + \left| d_{\sigma(j)} - g(U_{\sigma(j)})\right| > 3 L_1 \delta \;\Big|\; U_{\sigma(i)}, U_{\sigma(j)}\right)\\
&\le 2 \Pb\left( \left| d_{\sigma(i)} - g(U_{\sigma(i)}) \right| > \frac{3}{2} L_1 \delta \;\Big|\; U_{\sigma(i)}, U_{\sigma(j)} \right)\\
&\overset{(d)}{\le} 4\exp\left\{-2n^2 \left(\frac{3}{2}L_1\delta\right)^2\right\}
= 4\exp\left\{-\frac{9}{2}n^2 L_1^2 \delta^2\right\}.
\end{align*}
Here, $(d)$ is due to Hoeffding. The inequality in $(c)$ holds with probability at least $1-4\exp\{-2n\delta^2\}$. Letting two events
\begin{align*}
\calE_1 &= \left\{ \left| d_{\sigma(i)} - d_{\sigma(j)} \right| > 6 L_1 \delta \;\Big|\; U_{\sigma(i)}, U_{\sigma(j)} \right\} \\
\calE_2 &= \left\{ \left| g(U_{\sigma(i)}) - g(U_{\sigma(j)}) \right| < 3 L_1 \delta \right\},
\end{align*}
and using the fact that
\begin{align*}
\Pb\left( \calE_1 \right)
&=  \Pb\left( \calE_1 \cap \calE_2 \right) + \Pb\left( \calE_1 \cap \calE_2^c \right) \\
&\le \Pb\left( \calE_1 \cap \calE_2 \right) + \Pb\left( \calE_2^c \right),
\end{align*}
then we have
\begin{align*}
&\Pb\left( \left| d_{\sigma(i)} - d_{\sigma(j)} \right| > 6 L_1 \delta \;\Big|\; U_{\sigma(i)}, U_{\sigma(j)}\right)\\
&\le 4\exp\left\{-\frac{9}{2}n^2 L_1^2 \delta^2\right\} + 4\exp\{-2n\delta^2\}
\le 8\exp\{-2n\delta^2\},
\end{align*}
when $n > \frac{4}{9L_1^2}$. Putting $\delta = \frac{1}{6L_1} \sqrt{\frac{\log n}{n}}$, we have
\begin{align*}
\Pb\left( \left| d_{\sigma(i)} - d_{\sigma(j)} \right| > \sqrt{\frac{\log n}{n}} \;\Big|\; U_{\sigma(i)}, U_{\sigma(j)}\right) \le 8e^{-\frac{1}{18L_1^2}\log n}.
\end{align*}

We next prove the converse. First, by inverse Lipschitz we have
\begin{align}
\left| \frac{\sigma(i)}{n} - \frac{\sigma(j)}{n} \right|
&\le \left| \frac{\sigma(i)}{n} - U_{\sigma(i)} \right| + \left| \frac{\sigma(j)}{n} - U_{\sigma(j)} \right| + \left| U_{\sigma(i)} - U_{\sigma(j)} \right| \notag \\
&\le \left| \frac{\sigma(i)}{n} - U_{\sigma(i)} \right| + \left| \frac{\sigma(j)}{n} - U_{\sigma(j)} \right| + \frac{1}{L_2}\left| g(U_{\sigma(i)}) - g(U_{\sigma(j)}) \right|.
\label{eq:thm,converse,eq1}
\end{align}
By Dvoretzky, we have $\Pb\left( \left| \frac{\sigma(i)}{n} - U_{\sigma(i)} \right| > \eta \right) \le 2\exp\{ -2n \eta^2\}$ for any $\eta > 0$. Putting $\eta = \sqrt{\frac{1}{2n} \log\left(\frac{2}{\alpha}\right)}$, then $\Pb\left( \left| \frac{\sigma(i)}{n} - U_{\sigma(i)} \right| > \sqrt{\frac{1}{2n} \log\left(\frac{2}{\alpha}\right)} \right) \le \alpha$. That is,
\begin{align}
\left| \frac{\sigma(i)}{n} - U_{\sigma(i)} \right| \le \sqrt{\frac{1}{2n} \log\left(\frac{2}{\alpha}\right)}
\label{eq:thm,converse,eq2}
\end{align}
with probability at least $1-\alpha$.

Next, we note that
\begin{align*}
\left| g(U_{\sigma(i)}) - g(U_{\sigma(j)}) \right|
\le \left| g(U_{\sigma(i)}) - d_{\sigma(i)} \right| + \left| g(U_{\sigma(j)}) - d_{\sigma(j)} \right| + \left| d_{\sigma(i)} - d_{\sigma(j)} \right|.
\end{align*}
By Hoeffding, we know $\Pb\left( \left| g(U_{\sigma(i)}) - d_{\sigma(i)} \right| > \delta \right) \le 2\exp\{ -2n^2 \delta\}$ for any $\delta > 0$. Putting $\delta = \sqrt{\frac{1}{2n^2} \log\left(\frac{2}{\alpha}\right)}$, then
\begin{align}
\left| g(U_{\sigma(i)}) - d_{\sigma(i)} \right| \le \sqrt{\frac{1}{2n^2} \log\left(\frac{2}{\alpha}\right)},
\label{eq:thm,converse,eq3}
\end{align}
with probability at least $1-\alpha$.

Substituting \eref{eq:thm,converse,eq2}, \eref{eq:thm,converse,eq3} and that $\left| d_{\sigma(i)} - d_{\sigma(j)} \right| < \epsilon$ with probability at least $1-\alpha$ into \eref{eq:thm,converse,eq1}, we have
\begin{align}
\left| \frac{\sigma(i)}{n} - \frac{\sigma(j)}{n} \right|  \le 2\sqrt{ \frac{1}{2n} \log\left(\frac{2}{\alpha}\right)} + \frac{2}{L_2}\sqrt{\frac{1}{2n^2}\log\left(\frac{2}{\alpha}\right)} + \frac{\epsilon}{L_2},
\label{eq:|di-dj| bound converse}
\end{align}
which holds with probability at least $(1-\alpha)^5$.

Putting $\alpha = 8e^{- \frac{1}{18 L_1^2} \log n}$, $\epsilon = \sqrt{\frac{\log n}{n}}$, and using the fact that
\begin{align*}
\log\left(\frac{2}{\alpha}\right) = \log\left(\frac{1}{4}\right) + \frac{\log n}{18L_1^2} \le \frac{\log n}{18L_1^2},
\end{align*}
we have
\begin{align*}
\left| \frac{\sigma(i)}{n} - \frac{\sigma(j)}{n} \right| \le \sqrt{\frac{\log n}{n}} \left( \frac{1}{3L_1} + \frac{1}{3L_1L_2 \sqrt{n}} + \frac{1}{L_2}\right),
\end{align*}
with probability at least $(1-8e^{- \frac{1}{18 L_1^2} \log n})^5 \approx 1 - 40e^{- \frac{1}{18 L_1^2} \log n}$ for large $n$.

\subsection{Proof of Lemma 2}
For clarity and notational simplicity we prove a continuous version of the lemma. First, we define $w^{step}: [0,1]^2 \rightarrow [0,1]$ as the continuous version of $H^w \otimes \vone_{h \times h}$. That is, we equally partition $[0,1]$ into $k$ sub-intervals with width $h/n$. Then, for any $(x,y)$ in the $(i,j)$th sub-interval $[i(h/n),\, (i+1)(h/n)] \times [j(h/n),\, (j+1)(h/n)]$, we let $w^{step}(x,y) = H^w_{i,j}$.

By assumption that $w$ is smooth, there exists $\zeta_i \in \left[\frac{i-1}{k}, \frac{i}{k}\right]$ and $\xi_j \in \left[\frac{j-1}{k}, \frac{j}{k}\right]$ such that $w^{step}(u,v) = w(\zeta_i,\xi_j)$, for $u \in \left[\frac{i-1}{k}, \frac{i}{k}\right]$, and $v \in \left[\frac{j-1}{k}, \frac{j}{k}\right]$. Therefore, the approximation error is bounded as
\begin{align*}
\| w - w^{step} \|_{2}^2
&= \sum_{i=1}^k \sum_{j=1}^k \int_{\frac{i-1}{k}}^{\frac{i}{k}} \int_{\frac{j-1}{k}}^{\frac{j}{k}} \left( w(u,v) - w^{step}(u,v) \right)^2 dvdu\\
&= \sum_{i=1}^k \sum_{j=1}^k \int_{\frac{i-1}{k}}^{\frac{i}{k}} \int_{\frac{j-1}{k}}^{\frac{j}{k}} \left( w(u,v) - w^{step}(\zeta_i,\xi_j) \right)^2 dvdu\\
&\le \left(\frac{1}{k^2}\right)^2 \sum_{i=1}^k \sum_{j=1}^k \int_{\frac{i-1}{k}}^{\frac{i}{k}} \int_{\frac{j-1}{k}}^{\frac{j}{k}}
\sup_{ \tiny{\begin{array}{c} u \in [\frac{i-1}{k},\frac{i}{k}] \\ v \in [\frac{j-1}{k},\frac{j}{k}] \end{array}}} \left| \nabla w(u,v) \right|^2    dvdu\\
&\le \frac{1}{k^2} \sup_{u \in [0,1], v \in [0,1]} \left| \nabla w(u,v) \right|^2.
\end{align*}
Therefore,
\begin{align}
\| w - w^{step} \|_{2} \le \frac{1}{k}  \sup_{u,v \in [0,1]} \left| \nabla w(u,v) \right|.
\label{eq:thm2,proof,w-wstep}
\end{align}

\subsection{Proof of Lemma 3}
First, by definition of $\Hhat$ and $H$, we have
\begin{align}
\E[ \|\Hhat - H\|_{2}^2 ]
= \E\left[ \sum_{i=1}^k \sum_{j=1}^k \left(\frac{1}{h^2} \sum_{i_1=1}^h \sum_{j_1=1}^h \left(\Ahat_{ih+i_1,jh+j_1} - A_{ih+i_1,jh+j_1}\right)\right)^2 \right].
\label{eq:thm,hist,proof,||H-Hhat||2}
\end{align}
To evaluate \eref{eq:thm,hist,proof,||H-Hhat||2}, it is clear that we have to estimate
\begin{align*}
\E\left[ (\Ahat_{ij} - A_{ij})^2  \right]  \quad\mbox{and}\quad \E\left[ \Ahat_{ij} - A_{ij}  \right]
\end{align*}
for all $i,j = 1,\ldots,k$. Let $w_{ij}$ be the true graphon and
\begin{align*}
\what_{ij} = w(U_{\widehat{\sigma}(i)}, U_{\widehat{\sigma}(j)})
\end{align*}
be the empirical graphon ordered by $\sigmahat(1),\ldots,\sigmahat(n)$. Then it holds that
\begin{align}
\E\left[ (\Ahat_{ij} - A_{ij})^2  \right]
= \E[ (\Ahat_{ij}-\what_{ij})^2 + (A_{ij} - w_{ij})^2 + (w_{ij}-\what_{ij})^2],
\label{eq:thm,hist,proof,Ahat-A}
\end{align}
because $\E[\Ahat_{ij}] = \what_{ij}$ and $\E[A_{ij}] = w_{ij}$.

To bound \eref{eq:thm,hist,proof,Ahat-A}, we first show that
\begin{align}
\E[(\Ahat_{ij}-\what_{ij})^2] &= \Var[\Ahat_{ij}] \le 1, \label{eq:thm,hist,proof,Var Ahat}\\
\E[(A_{ij}-w_{ij})^2]         &= \Var[A_{ij}] \le 1.    \label{eq:thm,hist,proof,Var A}
\end{align}
Next, we bound the term $(w_{ij}-\what_{ij})^2$ as
\begin{align}
(w_{ij}-\what_{ij})^2
&\overset{(a)}{=}    \left[w(U_{{\sigma}(i)},U_{{\sigma}(j)})-w(U_{\widehat{\sigma}(i)},U_{\widehat{\sigma}(j)}) \right]^2 \notag \\
&\overset{(b)}{\le}  \left[ L \left(|U_{\sigma(i)} - U_{\widehat{\sigma}(i)}| + |U_{\sigma(j)} - U_{\widehat{\sigma}(j)}|\right) \right]^2 \notag \\
&\overset{(c)}{\le}  4C^2 L^2 \frac{\log n}{n}\label{eq:thm,hist,proof,w-what},
\end{align}
where $C = \frac{1}{3L_1} + \frac{1}{3L_1L_2} + \frac{1}{L_2}$. Here, in $(a)$ we write $w_{ij} = w(U_{{\sigma}(i)},U_{{\sigma}(j)})$. Since $w$ is the true graphon, the permutation $\sigma$ is the identity operator: $\sigma(i) = i$ for all $i$. The inequality in $(b)$ holds because of the Lipschitz condition on $w$. The inequality in $(c)$ is due to \eref{eq:|di-dj| bound converse}. Substituting \eref{eq:thm,hist,proof,Var Ahat}, \eref{eq:thm,hist,proof,Var A} and \eref{eq:thm,hist,proof,w-what} into \eref{eq:thm,hist,proof,Ahat-A} yields
\begin{equation}
\E\left[ (\Ahat_{ij} - A_{ij})^2  \right]  \le 2 + 4C^2 L^2 \frac{\log n}{n}.
\label{eq:thm,hist,proof,Ahat-A,2,result}
\end{equation}

Similarly, $\E\left[ \Ahat_{ij} - A_{ij}  \right]$ can be bounded as
\begin{align}
\E\left[ \Ahat_{ij} - A_{ij}  \right]
&\le \E\left[ \Ahat_{ij} - \what_{ij}\right] + \E\left[ w_{ij} - A_{ij}  \right] + \left| \what_{ij} - w_{ij}\right| \notag \\
&\le 2C L \sqrt{\frac{\log n}{n}}.
\label{eq:thm,hist,proof,Ahat-A,1,result}
\end{align}

Going back to \eref{eq:thm,hist,proof,||H-Hhat||2}, we can show that
\begin{align*}
&\E\left[ \sum_{i=1}^k \sum_{j=1}^k \left(\frac{1}{h^2} \sum_{i_1=1}^h \sum_{j_1=1}^h \left(\Ahat_{ih+i_1,jh+j_1} - A_{ih+i_1,jh+j_1}\right)\right)^2 \right] \\
&\le  \sum_{i=1}^k \sum_{j=1}^k \frac{1}{h^4} \left( h^2 \left(2+4C^2 L^2 \frac{\log n}{n}\right) + \frac{h^2(h^2-1)}{2} \left(2C L \sqrt{\frac{\log n}{n}} \right)^2\right)\\
&\le \frac{k^2}{h^2} \left(2+4C^2 L^2 \frac{\log n}{n}\right) + k^2 \left(2C^2 L^2 {\frac{\log n}{n}} \right).
\end{align*}
Substituting $n = kh$, we have
\begin{equation}
\E[ \|\Hhat - H\|_{2}^2 ]  \le  \frac{k^4}{n^2} \left(2+4C^2 L^2 \frac{\log n}{n}\right) + k^2\left(4C^2 L^2 {\frac{\log n}{n}} \right).
\end{equation}

\subsection{Proof of Lemma 4}
By definitions of $H_{ij}$ and $H_{ij}^w$, it holds that
\begin{align*}
\E[ H_{ij} ] = \E\left[ \frac{1}{h^2} \sum_{i_1=1}^h \sum_{j_1=1}^h A_{ih+i_1, jh+j_1}\right]
&= \frac{1}{h^2} \sum_{i_1=1}^h \sum_{j_1=1}^h \E\left[ A_{ih+i_1, jh+j_1}\right]\\
&= \frac{1}{h^2} \sum_{i_1=1}^h \sum_{j_1=1}^h w_{ih+i_1, jh+j1}
= H_{ij}^w.
\end{align*}
Consequently, we can show that
\begin{align*}
\E\left[ (H_{ij} - H_{ij}^w )^2\right] = \E\left[ (H_{ij})^2 \right] - (H_{ij}^w)^2,
\end{align*}
and hence
\begin{align*}
\E\left[ H_{ij}^2\right]
&= \frac{1}{h^4} \Big( \sum_{i_1=1}^h\sum_{j_1=1}^h\sum_{i_2\not=i_1}\sum_{j_2\not=j_1} \E\left[ A_{ih+i_1,jh+j_1}A_{ih+i2,jh+j_2} \right] \\
&\quad + \sum_{i_1=1}^h\sum_{j1=1}^h \E\left[ A_{ih+i_1,jh+j_1}^2 \right] \Big)\\
&= \frac{1}{h^4} \Big( \sum_{i_1=1}^h\sum_{j_1=1}^h\sum_{i_2\not=i_1}\sum_{j_2\not=j_1} w_{ih+i_1,jh+j_1} w_{ih+i2,jh+j_2} \\
&\quad + \sum_{i_1=1}^h\sum_{j1=1}^h  w_{ih+i_1,jh+j_1} \Big)\\
&= (H_{ij}^w)^2 + \frac{1}{h^4}\sum_{i_1=1}^h\sum_{j_1=1}^h w_{ih+i_1,jh+j_1}(1-w_{ih+i_1,jh+j_1})\\
&\le (H_{ij}^w)^2 + \frac{1}{h^2}.
\end{align*}
Therefore,
\begin{align*}
\E[\| H - H^w\|_2^2]= \sum_{i=1}^k \sum_{j=1}^k \E[ (H_{ij} - H_{ij}^w)^2 ] \le \frac{k^2}{h^2} = \frac{k^4}{n^2}.
\end{align*}

\subsection{Proof of Theorem 3}
By the definition of MSE, we have
\begin{align}
\MSE &\defequal\, \frac{1}{n^2} \E[\| \what^{est} - w \|_2^2] \label{eq:thm,consistency,mse}\\
&= \frac{1}{n^2} \Big( \E[ h^2 \| \what^{tv} - H^w \|_2^2] + \E[\|H^w \otimes \vone_{h\times h} - w\|_2^2] \notag\\
&\quad + 2\E[(\what^{tv} - H^w)^T(H^w \otimes \vone_{h\times h} - w)] \Big).  \notag
\end{align}

The first term above can be bounded by Lemma 5:
\begin{align*}
\| \what^{tv} - H^w \|_2^2 \le \varepsilon^2,
\end{align*}
because by assumption $\|\nabla H^w - (\nabla H^w)_s\|_1 = 0$. Now, $\varepsilon$ can further be bounded by Lemma 3 and Lemma 4:
\begin{align}
\varepsilon^2 &\defequal \E[\|\eta + \rho\|_{2}^2] \notag \\
&\overset{(a)}{=} \E[ \|\Hhat - H\|_{2}^2 ]  + \E[ \|H - H^w \|_{\ell_2}^2], \notag \\
&\le \frac{k^4}{ n^2 } \left(2+4C^2 L^2 \frac{\log n}{n}\right) + k^2 \left(4C^2 L^2 {\frac{\log n}{n}} \right) + \frac{k^4}{n^2},
\label{eq:thm,consistency,term1}
\end{align}
where in $(a)$ we used the fact that $\E[H_{ij}] = H_{ij}^w$ so that $\E[\rho] = 0$. Therefore,
\begin{align*}
&\frac{1}{n^2} \E\left[ h^2\| \what^{tv} - H^w \|_{2}^2 \right]\\
&= \frac{k^2}{ n^2 } \left(2+4C^2 L^2 \frac{\log n}{n}\right) + \left(4C^2 L^2 {\frac{\log n}{n}} \right) + \frac{k^2}{n^2} \rightarrow 0
\end{align*}
as $n \rightarrow \infty$ and $k/n \rightarrow 0$.

The second term in \eref{eq:thm,consistency,mse} can be bounded by Lemma 2, which gives
\begin{align}
\frac{1}{n^2} \|H^w \otimes \vone_{h\times h} - w\|_2^2
&\le \frac{C'}{k^2n^2} \rightarrow 0 \label{eq:thm,consistency,term2}
\end{align}
as $n\rightarrow \infty$, where $C' = \sup |\nabla w|^2$.

Substituting \eref{eq:thm,consistency,term1} and \eref{eq:thm,consistency,term2} into \eref{eq:thm,consistency,mse} completes the proof.

\bibliographystyle{plainnat}
\bibliography{chan_airoldi_ref}

\end{document}